\documentclass[amsmath,amssymb,aps]
{revtex4-2}

\usepackage{graphicx}
\usepackage{dcolumn}
\usepackage{bm}
\usepackage{chemformula}
\usepackage{multirow}

\usepackage[lofdepth,lotdepth]{subfig}

\begin{document}

\title{The effect of \ch{B}-site alloying on the electronic and opto-electronic properties of \ch{RbPbI_3}: A DFT study}%

\author{Anupriya Nyayban}%
\email{$\rm{anupriya_rs@phy.nits.ac.in}$}
\author{Subhasis Panda}
\email{subhasis@phy.nits.ac.in}
\affiliation{Department of Physics\\National Institute of Technology Silchar, Assam, 788010, India }
\author{Avijit Chowdhury}
\email{avijitiacs@gmail.com}
\affiliation{Department of Physics\\National Institute of Technology Silchar, Assam, 788010, India \\ and\\Department of Condensed Matter Physics and Material Sciences, S.N. Bose National Centre for Basic Sciences, JD Block, Sector III, Salt Lake City,
Kolkata 700106, India}
\date{\today}%


\begin{abstract}
Divalent cations mixed lead halide perovskites with enhanced performances, high stabilities, and reduced toxicity are requisite to make persistent progress in perovskite solar cells. However, the mixing strategy is not reported extensively in search of a lead reduced structure. Herein, we report the structural, electronic and optical properties of \ch{RbPb_{1-x}M_{x}I_3} (where, \ch{M}=\{\ch{Sn},\ch{Ge}\} and \ch{x}=\{$0.25$, $0.50$, $0.75$\}) by alloying the \ch{B}-site with \ch{Sn} and \ch{Ge}, using the density functional theory. The formation enthalpy is estimated for all \ch{RbPb_{1-x}M_{x}I_3} (with \ch{x}= $0.25$, $0.50$, $0.75$), which confirms stability for all the structures. The energy bandgap and density of states (DOS) have been thoroughly investigated. The energy bandgap decreases with the increasing \ch{Sn}/\ch{Ge} contents, the lowest bandgap of $1.850$ eV is observed at $\ch{x}=0.50$ in the case of \ch{RbPb_{1-x}Ge_{x}I_3} systems. Further, the effective masses and the binding energy of excitons and spectroscopic limited maximum efficiency (SLME) are also estimated for all the mixed systems. The exciton type is observed to change from Mott-Wannier to Frenkel type with increasing the contents of both \ch{Sn} and \ch{Ge} at the \ch{B}-site. The maximum efficiency of $23$\% is achieved using an active layer containing an equal admixture of \ch{Sn}/\ch{Ge} and \ch{Pb}. The estimated parameters of both the mixed systems are consistent with the available literature of similar types.

\end{abstract}

\maketitle

\section{Introduction}
The burgeoning photo conversion efficiency (PCE) toward the Shockley–Queisser limit makes perovskites a game-changing material in photovoltaic technology over a short time frame. The perovskite materials possess a few notable features, such as suitable energy band edge positions \cite{ref4}, high mobilities of photogenerated charge carriers \cite{ref5}, the lower binding energy of excitons \cite{ref6}, etc., which are perfect for solar cell applications \cite{ref7,ref8,ref9,ref10,ref11,ref12,ref13}. The rapid progress of perovskite materials inspires researchers to look at ways to make a more stable and less hazardous \ch{ABX_3} structure, where \ch{A}, \ch{B}, and \ch{X} are the organic or inorganic monovalent cations, inorganic divalent cations, and halides, respectively. Despite their great success, perovskites are still suffering from a few drawbacks, e.g., the device instability due to heat and moisture \cite{ref14,ref15}, the toxicity due to \ch{Pb}, etc. Therefore, further attempts to resolve these issues are required to shape the solar cell research based on perovskites materials. The inorganic halide perovskites have emerged as one of the hotspots in perovskite photovoltaics due to their higher thermal stability as compared to the organic-inorganic hybrid perovskites \cite{ref16,ref17,ref18}. Among the inorganic perovskites, \ch{CsPbX_3} was studied earlier \cite{ref19,ref20,ref21,ref22} and found to exhibit the most promising photovoltaic properties. In another study \cite{ref23}, \ch{CsSnI_3} is reported to have a direct energy bandgap ($1.3$ eV) and mobility as $400$ $\rm{cm^{-2}/Vs}$. However, very few studies are available for \ch{Rb}-based inorganic perovskites \cite{ref46,ref47,ref33,ref41}.

Despite their exceptional features and application scopes, several concerns, such as stability and toxicity, continue to hamper their performance at the device level. Hence, alloying the \ch{B}-site with suitable group materials, such as \ch{Sn} and \ch{Ge}, is a popular trend to deal with the toxicity issue of perovskites without compromising the performance \cite{ref24,ref25}. Both the materials pose the same oxidation state and structure, which facilitates the possible partial substitution of \ch{Pb} in \ch{MAPbI_3} \cite{ref24,ref25,ref26,ref27,ref28}. Better photoconversion efficiency and stability have been observed experimentally for the partial \ch{Sn} substitution in the \ch{Pb} site of the \ch{MAPbI_3} \cite{ref24,ref25}. It has been theoretically investigated \cite{ref28} and reported that the partial \ch{Ge} substituted \ch{MAPbI_3}, i.e., \ch{MAGe_{x}Pb_{1-x}I_3}, exhibits better photovoltaic properties and higher absorption. Ming-Gang Ju et al. \cite{ref31} have predicted that the energy bandgap and the absorption of \ch{RbSn_{0.5}Ge_{0.5}I_3} are suitable for PSC and its performance is comparable to that of \ch{MAPbI_3}. \ch{B}-site doping in \ch{CsPbI_3} with $\ch{Sn}^{2+}$ are also reported \cite{ref53} theoretically to improve the bandgap and the structural stability. \ch{CsPb_{0.7}Sn_{0.3}I_3} based solar cell is experimentally reported \cite{ref52} to achieve a PCE of $9.41$ \% and high $J_{sc}$ (short circuit current density) of $20.96$ $\rm{mAcm^{-2}}$.  Recently, the cubic phase of \ch{CsPb_{0.8}Ge_{0.2}I_3} are experimentally \cite{ref51} formed at $90^{\circ}$C, showing PCE of $3.97$ \%, external quantum efficiency of $40-70$ \% and poses stability at room temperature. To the best of our knowledge, no literature describes the partial substitution of either \ch{Sn} or \ch{Ge} in the divalent cation-site of \ch{RbPbI_3}. Therefore, in this work, we have partially substituted \ch{Sn} or \ch{Ge} in the divalent cation-site of \ch{RbPbI_3} and theoretically studied their electronic and optoelectronic properties using density functional theory. The structural, electronic, and optical properties are systematically investigated for both \ch{RbPb_{1-x}Sn_{x}I_3} and \ch{RbPb_{1-x}Ge_{x}I_3} with $\rm{\ch{x} = 0.25}$, $0.50$  and $0.75$. The computational details are described in Section II. The structural properties, lattice parameters, and formation energies are calculated and described in Section III-A. The energy band structure and density of states (DOS) under the electronic properties are investigated in detail and discussed in Section III-B. The optical properties, e.g., the imaginary part of the dielectric function and the absorption spectra for all the mixed systems, are described in Section III-C. Furthermore, the effective masses, exciton binding energies, and the spectroscopic limited maximum efficiencies are also studied and sectioned in III-C, followed by a brief conclusion in section IV.

\section{Computational methods}
All the first principle based calculations for $2 \times 2 \times 2$ supercell of both \ch{RbPb_{1-x}Sn_{x}I_3} and \ch{RbPb_{1-x}Ge_{x}I_3} (where $\ch{x} = 0.25$, $0.50$ and $0.75$) are performed using WIEN2k \cite{ref32} within the full potential linearized augmented plane wave (FP-LAPW) method. The supercell is created from the optimized orthorohmbic \ch{RbPbI_3} \cite{ref33} of \ch{NH_{4}CdCl_{3}} type structure having $Pnma$ space group. Each supercell is relaxed and optimized. The Muffin tin radius of $2.50$ {\AA} is set for all the atoms in all the mixed systems.  $\rm{RK_{max}}$ of $8$ is considered for both \ch{RbPb_{0.50}Sn_{0.50}I_3} and \ch{RbPb_{0.50}Ge_{0.50}I_3} whereas it is set to $7$ for all other systems. PBE (Perdew-Burke-Ernzerhof)-GGA (generalized gradient approximation) \cite{ref34} is used to treat the exchange correlation functional. Spin orbit coupling (SOC) effect is not  included here due to the fact that PBE without SOC estimate the bandgap accurately for the hybrid perovskites \cite{ref35,ref36,ref37,ref38} and even PBEsol \cite{ref39,ref40} and hybrid functional \cite{ref35} are reported to overestimate the bandgap. Moreover, PBE with SOC does not change the band structure pattern for both \ch{RbSnI_3} and \ch{RbGeI_3} \cite{ref41}, rather it reduces the bandgap. The electronic structure is calculated over the kmesh of $1 \times 6 \times 13$ and $8 \times 7 \times 2$ for $\ch{x}=0.50$ and others in \ch{RbPb_{1-x}Sn_{x}I_3} systems. In case of \ch{RbPb_{1-x}Ge_{x}I_3}, kmesh of $1 \times 5 \times 11$ and $7 \times 3 \times 3$ are set to calculate the electronic structure for $\ch{x}=0.50$ and all other values of $\ch{x}$. TB-mBJ (Tran-Blaha modified Becke-Johnson) \cite{ref42} potential is reported to estimate the bandgap accurately \cite{ref48,ref49} with less computation as compared to that with the hybrid functionals. Therefore, TB-mBJ potential is considered to find the bandgap for all the treated systems. Later, the higher kmesh of $15 \times 14 \times 4$, $3 \times 11 \times 25$ and $15 \times 7 \times 8$ are set to evaluate all the optical properties for $\ch{x}=\{0.25, 0.75\}$ in \ch{RbPb_{1-x}Sn_{x}I_3}, $\ch{x}=0.50$ in both the mixed cases and $\ch{x}=\{0.25, 0.75\}$ in \ch{RbPb_{1-x}Ge_{x}I_3}, respectively.  
 
\section{Results and observations}

\subsection{Structural and electronic properties}
The optimized lattice parameters for all the supercells, obtained using the Birch-Murnaghan equation of state \cite{ref44}, are represented in TABLE \ref{tab:latsnge}. The variation of total energy with volume fitted to the second order Birch-Murnaghan equation of state for all are represented in FIG. S1 in the Supporting Information. 

\begin{table}[h!]
\caption{Lattice parameters for \ch{RbPb_{1-x}M_{x}I_3} systems} 
\label{tab:latsnge}
\begin{tabular}{ccccccccc}
\hline
\hline
 \ch{M} & \ch{x} & a (\AA) & b (\AA) & c (\AA) & $\alpha$ & $\beta$ & $\gamma$ & $H$ (eV)\\
\hline
\multirow{ 3}{*}{\ch{Sn}} &  $0.25$ & $9.785$ & $10.530$ & $35.601$ & $90.000^{\circ}$ & $90.000^{\circ}$ & $90.000^{\circ}$ & $-1.932$\\
& $0.50$ & $35.478$ & $10.494$ & $4.876$ & $90.000^{\circ}$ &  $90.000^{\circ}$ & $90.000^{\circ}$ & $-3.312$\\
& $0.75$ & $9.784$ & $10.529$ & $35.598$ & $90.000^{\circ}$ & $90.000^{\circ}$ & $90.000^{\circ}$ & $-3.070$\\
\hline
\multirow{ 3}{*}{\ch{Ge}} & $0.25$ & $9.702$ & $20.883$ & $21.074$ & $119.701^{\circ}$ & $103.309^{\circ}$ & $90.000^{\circ}$ & $-1.562$\\
& $0.50$ & $34.943$ & $10.336$ & $4.802$ & $90.000^{\circ}$ & $90.000^{\circ}$ & $90.000^{\circ}$ & $-2.448$\\
& $0.75$ & $9.496$ & $20.438$ & $20.625$ & $119.701^{\circ}$ & $103.309^{\circ}$ & $90.000^{\circ}$ & $-1.922$\\
\hline
\hline
\end{tabular}
\end{table}

All the optimized supercells are also depicted  in FIG. S2 in the supporting information. TABLE \ref{tab:latsnge} also suggests that there is a decreasing trend in the lattice parameters for both \ch{RbPb_{1-x}Sn_{x}I_3} and \ch{RbPb_{1-x}Ge_{x}I_3} when the concentration of \ch{Sn} and \ch{Ge} increases, respectively. This is attributed to the smaller ionic radii of \ch{Sn} and \ch{Ge} as compared to that of \ch{Pb}. The volume of the supercell also decreases more with the increase of \ch{Ge} concentration than that of \ch{Sn} owing to the smaller \ch{Ge}-ionic radii. The lowest volume is observed for the equal mixture of both \ch{Pb}-\ch{Sn} and \ch{Pb}-\ch{Ge} cases. A material can be destabilized by the external effects e.g. heat, oxygen, and moisture. Therefore, it is necessary to find the enthalpy of the formation, which plays a significant role in determining the chemical as well as the thermodynamic stability of the material. A \ch{ABX_3}-type structure generally decomposes into \ch{AX} and \ch{BX_2}. Hence, the formation energy ($H$) for \ch{Pb}-\ch{Sn} and \ch{Pb}-\ch{Ge} mixing systems are calculated using the following reactions:
\begin{eqnarray}
\label{eq:fps}
H_{Sn}=E(\ch{RbPb_{1-x}Sn_{x}I_3}) - E(\ch{RbI}) - \ch{x} E(\ch{SnI_2}) -2(1-\ch{x}) E(\ch{I}) -(1-\ch{x}) E(\ch{Pb}) \\
H_{Ge}=E(\ch{RbPb_{1-x}Ge_{x}I_3}) - E(\ch{RbI}) - \ch{x} E(\ch{GeI_2}) -2 (1-\ch{x}) E(\ch{I}) - (1-\ch{x}) E(\ch{Pb})
\end{eqnarray}
where $\ch{x}$ and $E$ symbolize the concentration percentage of \ch{Sn}/\ch{Ge} and the total energy corresponding to each compound, respectively. The calculated total energy for \ch{RbPb_{1-x}Sn_{x}I_3}, \ch{RbPb_{1-x}Ge_{x}I_3}, \ch{SnI_2}, \ch{GeI_2}, \ch{RbI}, \ch{Pb} and \ch{I} are listed in TABLE SI in the Supporting Information. Further, the enthalpy of formation per formula unit for both the systems are also calculated using relation (1)-(2) and listed in TABLE \ref{tab:latsnge}. It is observed that the negative values of the enthalpy of formation is increasing with the concentration of \ch{Sn} and \ch{Ge} excepts the highest enthalpies are for $\ch{x}=0.50$ for both \ch{RbPb_{1-x}Sn_{x}I_3} and \ch{RbPb_{1-x}Ge_{x}I_3}. These values suggest the stability of all the structures whereas the highest stability is observed at $\ch{x}=0.50$. Additionally, the formation energy can be explained by the bond strength of \ch{B}-\ch{I}. The stronger bond again is determined from the smaller bond length and the higher electronegativity difference between \ch{B} and \ch{I}. The difference of electronegativities between \ch{Pb}-\ch{I}, \ch{Sn}-\ch{I} and \ch{Ge}-\ch{I} are $0.33$, $0.70$ and $0.56$ respectively. The higher electronegativity difference for \ch{Sn}-\ch{I} (for \ch{RbPb_{1-x}Sn_{x}I_3}) indicates the stronger bonds and the higher enthalpies as compared with \ch{RbPb_{1-x}Ge_{x}I_3}. Therefore, the calculated enthalpy of formation and \ch{B}-\ch{I} bond strength both follow the condition of a stable structure.

\subsection{Electronic Properties}
The electronic bandgap values are crucial for an absorber used in solar cell owing to the fact that it makes a solar cell suitable to absorb photons and achieve maximum efficiency with minimum optical losses. Therefore, the calculated values of the bandgaps for all are listed in TABLE \ref{tab:bgpsg}. 
\begin{table}[h!]
\caption{The bandgap $E_g$ in eV for all \ch{Pb}-\ch{Sn} and \ch{Pb}-\ch{Ge} mixed systems}
\label{tab:bgpsg}
\begin{tabular}{ccccccc}
\hline
\hline
& \multicolumn{3}{c}{\ch{RbPb_{1-x}Sn_{x}I_3}} & \multicolumn{3}{c}{\ch{RbPb_{1-x}Ge_{x}I_3}}\\
\hline
 & \ch{x}=0.25 & \ch{x}=0.50 & \ch{x}=0.75 & \ch{x}=0.25 & \ch{x}=0.50 & \ch{x}=0.75 \\
\hline
PBE & $2.056$ & $1.960$ & $1.951$ & $2.289$ & $1.850$ & $2.153
$\\
TB-mBJ & $2.450$ & $2.464$ & $2.434$ & $2.757$ & $2.276$ & $2.581$ \\
\hline
\hline
\end{tabular}
\end{table}

\begin{figure}[h!]
\centering
  \subfloat[\ch{RbPb_{0.75}Sn_{0.25}I_3}.]{\includegraphics[scale=0.5]{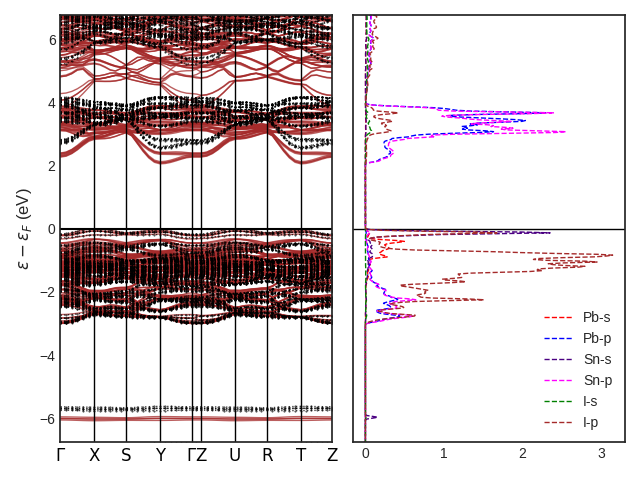}\label{fig:bdsn25}}
  \subfloat[\ch{RbPb_{0.25}Sn_{0.75}I_3}.]{\includegraphics[scale=0.5]{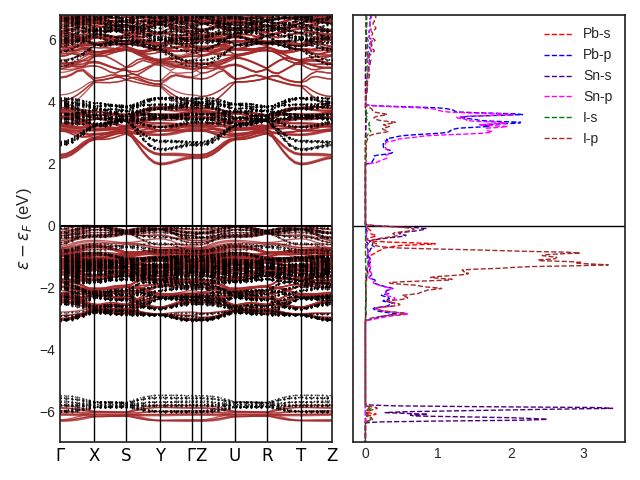}\label{fig:bdsn75}}
    \hspace{0.1cm}
  \subfloat[\ch{RbPb_{0.50}Sn_{0.50}I_3}.]{\includegraphics[scale=0.5]{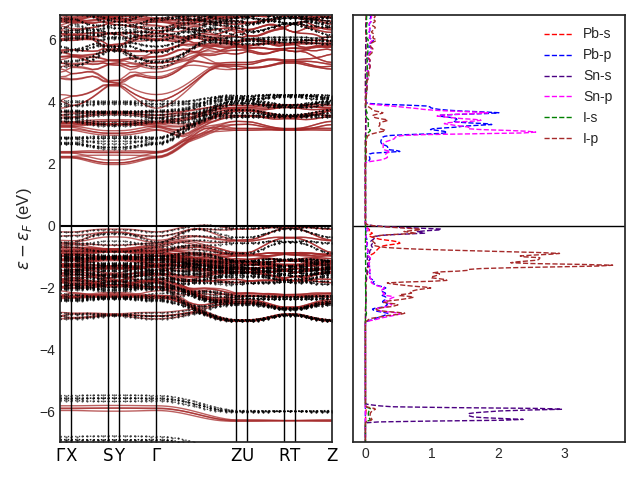}\label{fig:bdsn50}}
  \subfloat[\ch{RbPb_{0.50}Ge_{0.50}I_3}.]{\includegraphics[scale=0.5]{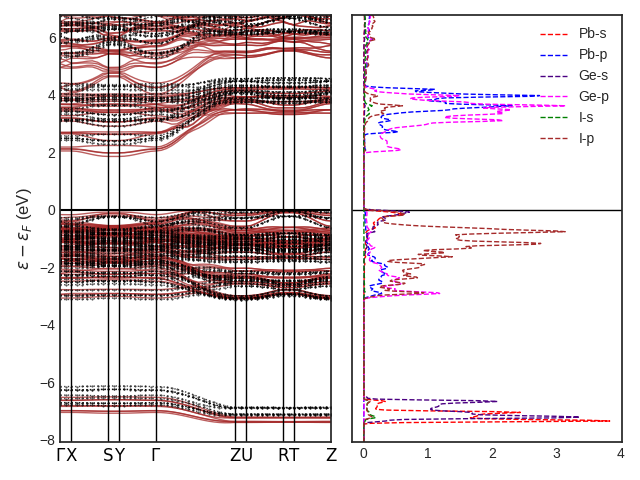}\label{fig:bdge50}}    
   \caption{The bandstructure (left side) and PDOS (right side) for \ref{fig:bdsn25}) \ch{RbPb_{0.75}Sn_{0.25}I_3}, \ref{fig:bdsn75}) \ch{RbPb_{0.25}Sn_{0.75}I_3}, \ref{fig:bdsn50}) \ch{RbPb_{0.50}Sn_{0.50}I_3}, \ref{fig:bdge50} \ch{RbPb_{0.50}Ge_{0.50}I_3}, respectively. The solid and dotted line in the bandstructure represents the bandstructure calculated with PBE and TB-mBJ, respectively.}\label{fig:bdpsg1}
\end{figure}

\begin{figure}[h!]
\centering
  \subfloat[\ch{RbPb_{0.75}Ge_{0.25}I_3}.]{\includegraphics[scale=0.5]{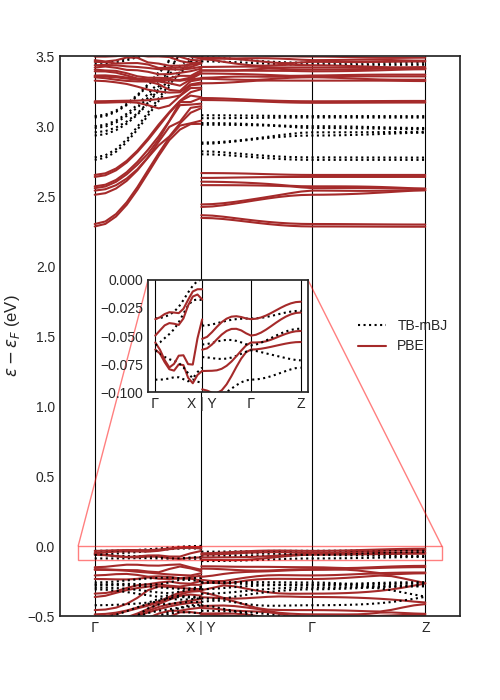}\label{fig:bdge25}}
  \subfloat[\ch{RbPb_{1-x}Ge_{x}I_3}.]{\includegraphics[scale=0.5]{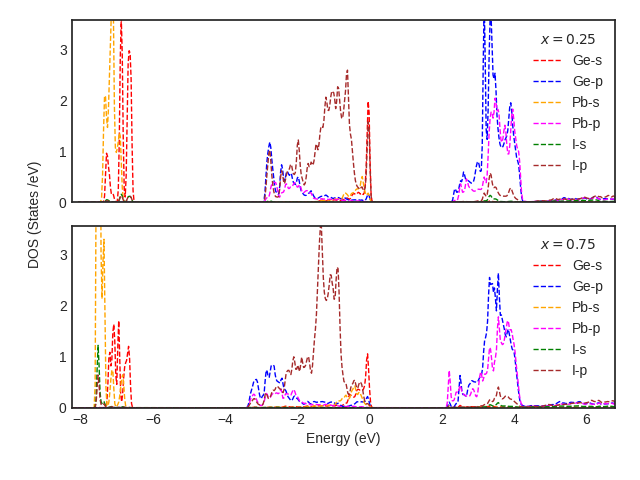}\label{fig:dg27}}
    \hspace{0.1cm} 
   \subfloat[\ch{RbPb_{0.25}Ge_{0.75}I_3}.]{\includegraphics[scale=0.5]{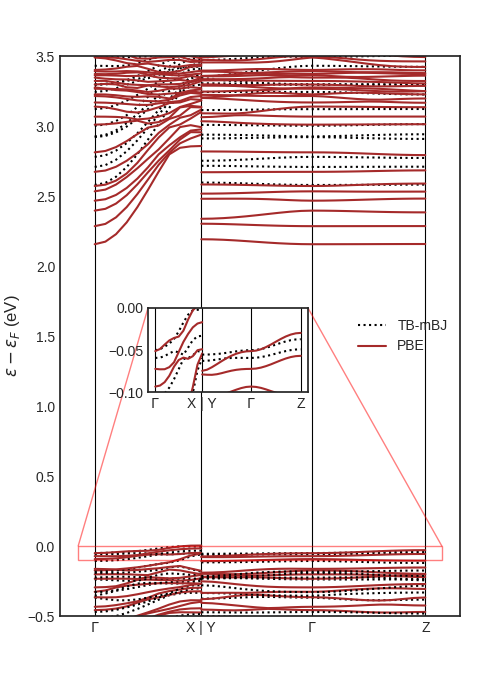}\label{fig:bdge75}} 
   \caption{The bandstructure of (\ref{fig:bdge25}) \ch{RbPb_{0.75}Ge_{0.25}I_3}, (\ref{fig:bdge75}) \ch{RbPb_{0.25}Ge_{0.75}I_3} and (\ref{fig:dg27}) the partial density of states for \ch{RbPb_{1-x}Ge_{x}I_3}.}\label{fig:bdpsg2}
\end{figure}

The bandgap values calculated with PBE decrease gradually with the increase of \ch{Sn} concentration for \ch{Pb}-\ch{Sn} mixed systems. 
 The Fermi level is fixed to zero for all the structures. The DOS estimated with PBE and the bandstructures estimated with both PBE and TB-mBJ potentials are shown for \ch{Pb}-\ch{Sn} mixed systems and \ch{RbPb_{0.5}Ge_{0.5}I_3} in FIG. \ref{fig:bdpsg1} while it is shown in FIG. \ref{fig:bdpsg2} for \ch{RbPb_{0.75}Ge_{0.25}I_3} and \ch{RbPb_{0.25}Ge_{0.75}I_3}.    

The FIG. \ref{fig:bdpsg1} suggests that the conduction band minimum (CBM) and the valence band maximum (VBM) are observed at $\rm{Y}$, $\rm{S}$, $\rm{T}$ and $\rm{U}$, $\rm{Z}$, $\rm{X}$ for \ch{RbPb_{1-x}Sn_{x}I_3} when $\ch{x}$ values are $0.25$, $0.50$ and $0.75$, respectively. The CBM are observed at $\rm{Z}$, $\rm{Y}$, $\rm{\Gamma}$ while the VBM are found at $\rm{X}$, $\rm{Z}$ high symmetry $k$-points for \ch{RbPb_{1-x}Ge_{x}I_3} when \ch{Ge} concentration increases from $0.25$, $0.50$ to $0.75$, respectively. This confirms that all the structures have the indirect bandgaps. The bandgap does not decrease linearly (Vegard's law) with $\ch{x}$ rather it follows an anomalous behavior for both \ch{RbPb_{1-x}Sn_{x}I_3} and \ch{RbPb_{1-x}Ge_{x}I_3}. The similar behavior is also observed for \ch{MAPb_{x}Sn_{1-x}I_3} \cite{ref45}. This is owing to the variation of the symmetry for both \ch{RbPb_{1-x}Sn_{x}I_3} and \ch{RbPb_{1-x}Ge_{x}I_3} structures with the changing $\ch{x}$ values. The calculated bandgap values can not be compared with any experimental values due to the unavailability of the reported data in the literature. Although PBE without SOC are reported \cite{ref35,ref36,ref37,ref38} to estimate the bandgaps accurately, more advanced TB-mBJ potential is also used to estimate the band structures. TB-mBJ is reported \cite{ref49} to estimate the bandgap for the hybrid organic-inorganic perovskites efficiently with the affordable computational cost as compared to that with both hybrid functionals and GW method. The TB-mBJ estimated band structures show that the bandgaps are increased by $0.394$, $0.504$, $0.483$ eV as compared to that with PBE for $\ch{x}=0.25, 0.50$ and $0.75$, respectively in the case of \ch{Pb}-\ch{Sn} mixed systems. In case of \ch{RbPb_{1-x}Ge_{x}I_3}, the bandgaps are widened by $0.468$, $0.426$ and $0.431$ eV for \ch{Ge} concentrations of $0.25$, $0.50$ and $0.75$, respectively. The CBM are mainly shifted to the higher energy region whereas no significant changes are observed in the VBM for all the structures. The similar nature of bandstructures calculated with PBE and TB-mBJ potentials are also observed for pristine \ch{RbPbI_3} \cite{ref33}, \ch{RbSnI_3} and \ch{RbGeI_3} \cite{ref41}. 

The density of states (DOS) are also estimated  to gain a brief understanding of the behavior of atomic orbitals into the electronic structures. PDOS for \ch{RbPb_{1-x}Sn_{x}I_3} (where, $\ch{x}$ = $0.25$, $0.50$ and $0.75$) and \ch{RbPb_{0.50}Ge_{0.50}I_3} are plotted to the right hand column in FIG. \ref{fig:bdpsg1} whereas FIG. \ref{fig:bdge75} represents the PDOS for both \ch{RbPb_{0.75}Ge_{0.25}I_3} and \ch{RbPb_{0.25}Ge_{0.75}I_3}. The orbitals of $\ch{Rb}^+$ cation contribute to the energy region far below the bandgap region and they are found to be well localized. Hence, \ch{Rb} atom does not contribute to the the active region of both the structures. Thus, the contribution of \ch{Pb}, \ch{Sn}, \ch{I} orbitals for \ch{RbPb_{1-x}Sn_{x}I_3} and the orbitals of \ch{Pb}, \ch{Ge}, \ch{I} in case of \ch{RbPb_{1-x}Ge_{x}I_3} are represented and discussed to get a clear understanding near the bandgap. PDOS plot suggests that the uppermost region of the first valence band is dominated by the \ch{I}-5p and \ch{Pb}-6s orbitals. Since \ch{Sn} and \ch{Ge} are of similar electronic structures as \ch{Pb}, it also shows a similar behavior in the uppermost region of the first VB for both \ch{RbPb_{1-x}Sn_{x}I_3} and \ch{RbPb_{1-x}Ge_{x}I_3}. The lower region of the first VB is mainly contributed by \ch{I}-5p with a minor population of \ch{Sn}-5p and \ch{Pb}-6p in case of \ch{RbPb_{1-x}Sn_{x}I_3} whereas \ch{Ge}-4p and \ch{Pb}-6p for \ch{RbPb_{1-x}Ge_{x}I_3}. The first VB energy levels are extended from $0$ to $-3.063$, $-3.158$ and $-3.114$ eV with the increasing $\ch{x}$ values ($0.25$, $0.50$ and $0.75$) for \ch{RbPb_{1-x}Sn_{x}I_3}. In case of \ch{RbPb_{1-x}Ge_{x}I_3}, the first VB ranges from $0$ to $-2.954$, $-3.152$ and $-3.433$ eV when \ch{Ge} concentration changes to $0.25$, $0.50$ and $0.75$, respectively. The more or less similar contribution of \ch{I}-5p are observed for all the systems. The contribution  of \ch{Sn}-5s (\ch{Ge}-4s) in the first VB region is decreasing when the concentration of \ch{Sn} (\ch{Ge}) increases. The second VB are dominated by \ch{Sn}-5s and it increases with the increase of \ch{Sn} content in \ch{RbPb_{1-x}Sn_{x}I_3}. Similarly, the contribution of \ch{Pb}-6s and \ch{Ge}-4s are observed for the second VB where \ch{Pb}-6s is dominant and the contribution of \ch{Ge}-4s
is increased with the increasing \ch{Ge} concentration. Again, the conduction band is predominantly occupied by \ch{Pb}-6p and \ch{Sn}-5p (\ch{Ge}-4p) orbitals for \ch{RbPb_{1-x}Sn_{x}I_3} (\ch{RbPb_{1-x}Ge_{x}I_3}). The contribution of \ch{Sn}-5p and \ch{Ge}-4p to the CB increases with the increasing \ch{Sn} and \ch{Ge} content for the \ch{Sn} and \ch{Ge} mixed systems, respectively. The upper region of CB also shows a minor contribution of \ch{I}-5s and \ch{I}-5p orbitals for all the mixed systems under consideration. The CB energy levels ranges from $2.025$ to $3.957$, $1.957$ to $3.944$ and $1.920$ to $3.906$ eV when \ch{Sn} concentrations are $0.25$, $0.50$ and $0.75$, respectively for \ch{Sn} mixed systems. Similarly, the CB in \ch{RbPb_{1-x}Ge_{x}I_3} extends from $2.119$ to $4.295$, $1.855$ to $4.304$ and $2.243$ to $4.310$ eV with the \ch{Ge} concentration of $0.25$, $0.50$ and $0.75$, respectively. The similar behavior are also reported \cite{ref28} for the mixed \ch{MAGe_{x}Pb_{1-x}I_3} systems. Therefore, the hybrid states of $\ch{Pb}^{2+}$ and $\ch{Sn}^{2+}$ ($\ch{Ge}^{2+}$) mainly determine the photovoltaic properties whereas it is determined by $\ch{Pb}^{2+}$ states in pristine \ch{RbPbI_3} \cite{ref33}. 

\begin{figure}[h!]
\centering
  \subfloat[\ch{RbPb_{1-x}Sn_{x}I_3}.]{\includegraphics[scale=0.5]{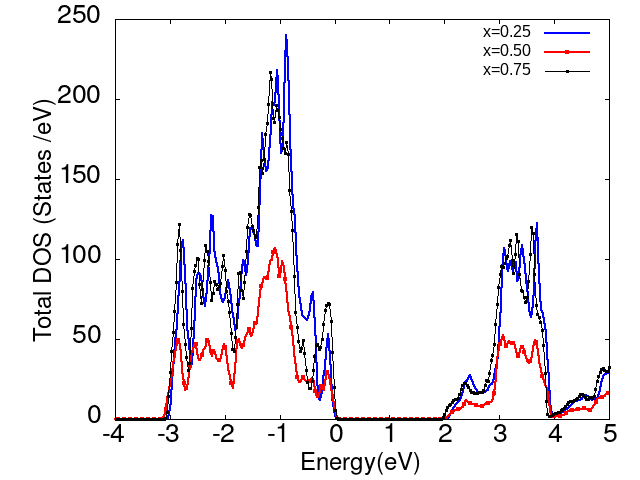}\label{fig:tds}}
  \subfloat[\ch{RbPb_{1-x}Ge_{x}I_3}.]{\includegraphics[scale=0.5]{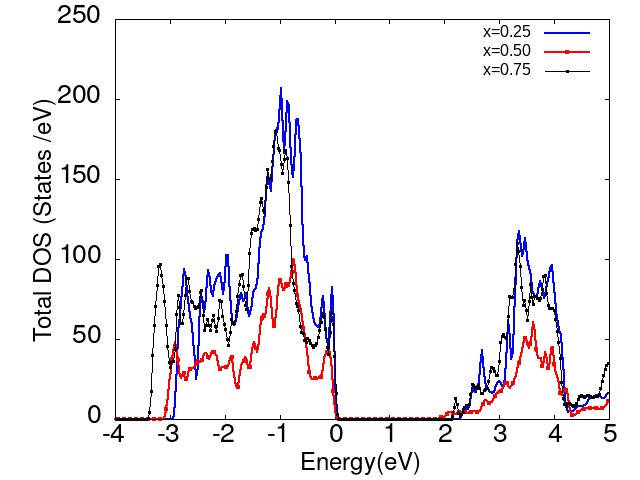}\label{fig:tdg}}
   \caption{The total density of states for both the mixing cases.}\label{fig:tdpsg2}
\end{figure}

The total density of states (TDOS) for $\ch{x}$ =$0.25$, $0.50$ and $0.75$ are plotted in FIG. \ref{fig:tds} and FIG. \ref{fig:tdg} for the all \ch{Sn} and \ch{Ge} mixed systems. The figure shows the rise in VB edge is more compared to that of CB edge for all the mixed systems whereas the rise in the VB edge is highest for \ch{Sn} mixed systems. The calculated TDOS also indicates the increase in the no of states are more for \ch{RbPb_{0.75}Sn_{0.25}I_3} and \ch{RbPb_{0.25}Sn_{0.75}I_3} compared to \ch{RbPb_{0.50}Sn_{0.50}I_3}. Similar behavior is observed for \ch{Ge} mixed systems where the lesser no of states are found for $\ch{x}=0.50$. Hence, a large no of states are observed for the smaller and higher concentration of \ch{Pb} for all the mixed systems. Therefore, the probability of higher carrier concentration is observed at $\ch{x}=0.25$ and $\ch{x}=0.75$ for both the mixed systems and this can enhance the transport as well as the device properties. 

\subsection{Optical Properties}
\begin{figure}[h!]
\centering
  \subfloat[The imaginary part of the dielectric function for \ch{RbPb_{1-x}Sn_{x}I_3}.]{\includegraphics[width=6cm,height=7cm]{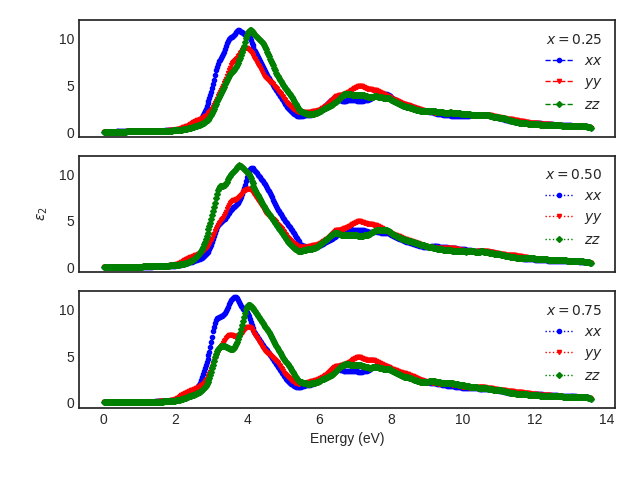}\label{fig:e2sn}}
  \subfloat[The absorption coefficients for \ch{RbPb_{1-x}Sn_{x}I_3}.]{\includegraphics[width=6cm,height=7cm]{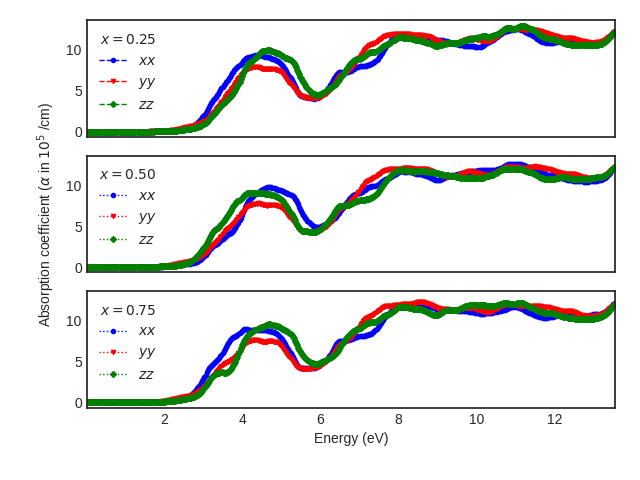}\label{fig:absn}}
  \hspace{0.1cm}
     \subfloat[The imaginary part of the dielectric function for \ch{RbPb_{1-x}Ge_{x}I_3}.]{\includegraphics[width=6cm,height=7cm]{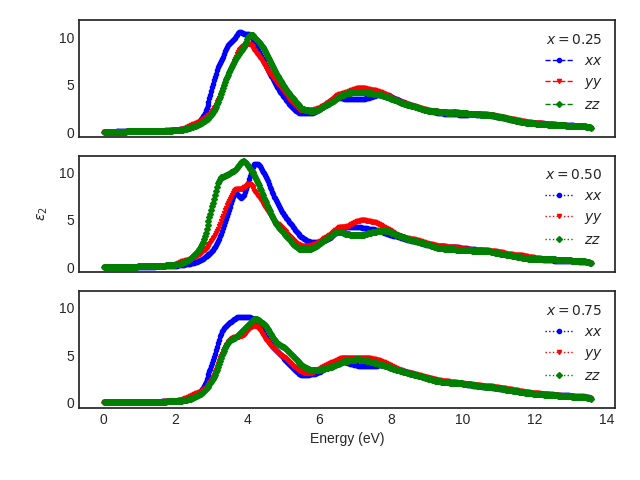}\label{fig:e2ge}}
      \subfloat[The absorption coefficients for \ch{RbPb_{1-x}Ge_{x}I_3}.]{\includegraphics[width=6cm,height=7cm]{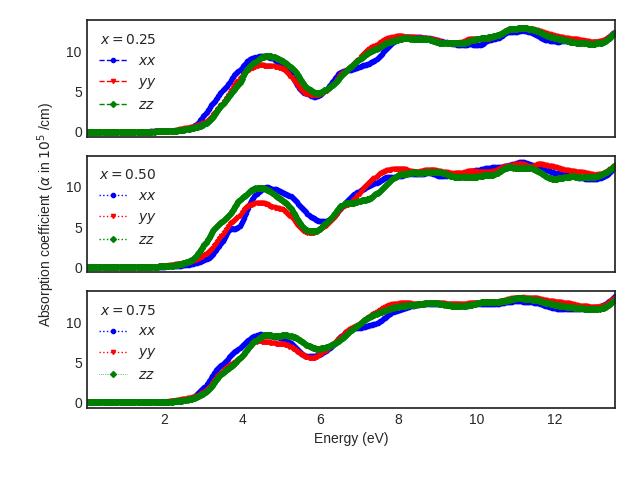}\label{fig:abge}}
   \caption{The variation of $\epsilon_{2}$ and $\alpha$ with the photon energy along three different polarization direction for $\ch{x}=0.25$, $0.50$ and $0.75$, respectively.}\label{fig:e1e2ab}
\end{figure}

The higher absorption and larger dielectric constant are crucial for the better performance of a solar cell. Therefore, the optical properties are estimated with the linear response method using the dielectric function $\epsilon(\omega) = \epsilon_1(\omega) + i \epsilon_2(\omega)$. The imaginary part of the dielectric function $\epsilon_2(\omega)$ is estimated using the momentum matrix involving the occupied and unoccupied states. FIG \ref{fig:e2sn} and \ref{fig:e2ge} represent $\epsilon_2(\omega)$ for \ch{RbPb_{1-x}Sn_{x}I_3} and \ch{RbPb_{1-x}Ge_{x}I_3}, respectively. The first and the strongest peaks are observed at $3.741$, $3.959$, $4.068$; $4.122$, $4.013$, $3.768$; and $3.659$, $4.013$, $4.040$ eV for \ch{RbPb_{0.75}Sn_{0.25}I_3}, \ch{RbPb_{0.50}Sn_{0.50}I_3} and \ch{RbPb_{0.25}Sn_{0.75}I_3} along x, y and z polarization directions whereas the second stronger peak is found near a photon energy of $7.000$ eV. On the other hand, the first strongest peaks are observed at the photon energy of $3.796$, $4.014$, $4.095$; $4.259$, $4.041$, $3.878$; $3.769$, $4.204$, $4.231$ along different directions (x, y abd z) eV when the \ch{Ge} concentration in \ch{RbPb_{1-x}Ge_{x}I_3} increases as $0.25$, $0.50$ and $0.75$, respectively. The bandgap reduction associated with the increase of \ch{Sn}/\ch{Ge} concentration (except at $\ch{x}=0.50$ for \ch{RbPb_{1-x}Ge_{x}I_3}) is responsible for such trend of peaks observed in all the mixed systems.  

The real part of the dielectric function $\epsilon_{1}(\omega)$ is estimated using $\epsilon_{2}(\omega)$ and the Kramer-Kronig relation. The variation of $\epsilon_{1}(\omega)$ with the photon energy for all the mixed systems are shown in FIG. S3a and S3b in the supporting information whereas the values of the static dielectric constants are provided in the TABLE \ref{tab:e1rn}. The average static dielectric constants ($\epsilon_{1}(0)$) are found to be $5.137$, $5.232$ and $5.300$  for \ch{RbPb_{1-x}Sn_{x}I_3} at the \ch{Sn}-concentrations of $0.25$, $0.50$ and $0.75$, respectively, while these values are $5.174$, $5.377$ and $5.299$ for all the \ch{Ge} mixed systems. A large static dielectric constant induces the higher charge screening. This will reduce the charge defect levels which will again reduce the radiative electron hole recombination. Thus, the large values of $\epsilon_{1}(0)$ for all the mixed systems will help the material to be used as an efficient photovoltaic absorber.   

\begin{table}[h!]
\caption{The static dielectric constant, reflectivity, refractive index, binding energy of excitons and the radius of the lowest bound states for all the systems.}
\label{tab:e1rn}
\begin{tabular}{ccccccccccccc}
\hline
\hline
\ch{M} & \ch{x} & \multicolumn{3}{c}{$\epsilon_{1}(0)$} & \multicolumn{3}{c}{$R(0)$} & \multicolumn{3}{c}{$n(0)$} & $E_{b}$ & $a^{*}$ \\
\hline
 & & $(100)$ & $(010)$ & $(001)$ & $(100)$ & $(010)$ & $(010)$ & $(100)$ & $(010)$ & $(010)$ & (meV) & \AA \\
 \hline
\multirow{ 3}{*}{\ch{Sn}} & $0.25$ & $5.408$ & $4.943$ & $5.061$ & $0.159$ & $0.144$ & $0.148$ & $2.326$ & $2.223$ & $2.250$ & $54.629$ & $25.637$ \\ 
 & $0.50$ & $5.143$ & $5.035$ & $5.519$ & $0.151$ & $0.147$ & $0.162$ & $2.268$ & $2.244$ & $2.349$ & $46.702$ & $29.444$ \\ 
 &  $0.75$ & $5.622$ & $5.093$ & $5.186$ & $0.165$ & $0.149$ & $0.152$ & $2.371$ & $2.257$ & $2.278$ & $333.101$ & $4.075$ \\ 
\hline
\multirow{ 3}{*}{\ch{Ge}} & $0.25$ & $5.436$ & $5.004$ & $5.081$ & $0.160$ & $0.146$ & $0.149$ & $2.332$ & $2.237$ & $2.254$ & $49.279$ & $28.217$ \\ 
 & $0.50$ & $5.177$ & $5.143$ & $5.810$ & $0.152$ & $0.151$ & $0.171$ & $2.275$ & $2.268$ & $2.410$ & $57.858$ & $23.125$ \\ 
 &  $0.75$ & $5.486$ & $5.160$ & $5.252$ & $0.161$ & $0.151$ & $0.154$ & $2.342$ & $2.272$ & $2.292$ & $96.384$ & $14.086$ \\
\hline
\hline
\end{tabular}
\end{table}

Moreover, all other optical properties such as extinction coefficient ($k(\omega)$), absorption coefficient ($\alpha(\omega)$), reflectivity ($R(\omega)$) and refractive index $n(\omega)$ are estimated using the following formulae:

\begin{eqnarray}
k(\omega)&=&\frac{\sqrt{\sqrt{{\epsilon_{1}(\omega)}^{2}+{\epsilon_{2}(\omega)}^{2}}+\epsilon_{1}(\omega)}}{\sqrt{2}}\\
\alpha(\omega)& = &\frac{2 \omega k(\omega)}{c}\\
R(\omega)&=& \frac{(n(\omega)-1)^{2}+{k(\omega)}^{2}}{(n(\omega)+1)^{2}+{k(\omega)}^{2}}\\
n(\omega)&=&\frac{\sqrt{\sqrt{{\epsilon_{1}(\omega)}^{2}+{\epsilon_{2}(\omega)}^{2}}-\epsilon_{1}(\omega)}}{\sqrt{2}}
\end{eqnarray}

FIG. \ref{fig:absn} and \ref{fig:abge} show the behavior of the absorption coefficients at different energy for all the \ch{Sn} and \ch{Ge} mixed systems. The first strong absorption peaks of \ch{RbPb_{1-x}Sn_{x}I_3} are found at the photon energy of $4.422$, $4.340$ and $4.694$ eV for $\ch{x}=0.25$; $4.694$, $4.395$ and $4.340$ eV for $\ch{x}=0.50$; $4.068$, $4.340$ and $4.694$ eV for $\ch{x}=0.75$ along x, y, z directions, respectively. In case of \ch{Ge} mixed systems, the first strong absorption peaks are observed at $4.503$, $4.476$ and $4.667$; $4.640$, $4.390$ and $4.449$; $4.449$, $4.422$ and $4.640$ eV of photon energy along x, y and z polarization directions for the \ch{Ge}-concentrations of $0.25$, $0.50$ and $0.75$, respectively. The first strong peaks are appeared due to the electronic transitions from the mixed valence state of \ch{I}-5p, \ch{Pb}-6s and \ch{Sn}-5s to the mixed conduction state of \ch{Sn}-5p and \ch{Pb}-6p for \ch{RbPb_{1-x}Sn_{x}I_3} whereas it is due to the transition from the valence state consisting \ch{I}-5p, \ch{Pb}-6s and \ch{Ge}-4s to the conduction state occupied by \ch{Pb}-5p and \ch{Ge}-4p for the all the \ch{Ge} mixed states. It is difficult to analyze the interband transitions exactly corresponding to the critical point as the large number of bands are involved for the orthorhombic system compared to the cubic one. Yet, it is interestingly observed from the absorption spectra for all the mixed systems that the absorption regions are widened as compared to that of pristine \ch{RbPbI_3} case \cite{ref33} and this is expanded to both visible and ultraviolet regions. Therefore, \ch{Sn} and \ch{Ge} mixed systems show the ability to absorb the wide range of photon energy from the solar spectrum compared to the pristine \ch{RbPbI_3} \cite{ref33}. 

\begin{table}[h!]
\caption{The calculated average integrated absorption intensity in $10^{3}$ eV/cm for \ch{RbPb_{1-x}M_{x}I_3}}
\label{tab:iipsg}
\begin{tabular}{ccccccc}
\hline
\hline
& \multicolumn{3}{c}{$\ch{M}=\ch{Sn}$} & \multicolumn{3}{c}{$\ch{M}=\ch{Ge}$}\\
\hline
Energy range (eV) & $0-1.7$ & $1.7-3.3$ & $3.3-5.0$ & $0-1.7$ & $1.7-3.3$ & $3.3-5.0$ \\
\hline
$\ch{x}=0.25$ & $2.907$ & $112.347$ & $1224.86$ & $2.839$ & $103.477$ & $1283.39$\\
$\ch{x}=0.50$ & $3.051$ & $133.66$ & $1205.98$ & $3.073$ & $129.155$ & $1227.267$ \\
$\ch{x}=0.75$ & $3.267$ & $162.843$ & $1217.186$ & $2.823$ & $107.949$ & $1099.527$ \\
\hline
\hline
\end{tabular}
\end{table}

Moreover, the integrated absorption intensity are estimated using the integration over the absorption curves along different energy regions for all the mixed systems. TABLE \ref{tab:iipsg} shows the estimated integrated absorption intensities along the three most important energy regions of the solar spectrum  (near-infrared, visible and ultraviolet). The integrated intensities over the energy range of $0-1.7$ and $1.7-3.3$ eV for \ch{Sn} mixed systems increase with the increase of \ch{Sn} concentration while the highest integrated intensity in the energy range of $0-1.7$ and $1.7-3.3$ eV are observed for \ch{RbPb_{0.50}Ge_{0.50}I_3} in the case of all \ch{Ge} mixed systems. The near-infrared region is interesting due to the fact that $55$\% of the solar energy reaching the earth are of this region. \ch{RbPb_{0.25}Sn_{0.75}I_3} and \ch{RbPb_{0.50}Ge_{0.50}I_3} exhibit the highest absorption in the near infrared region. For all the \ch{Pb}-\ch{Sn} and \ch{Pb}-\ch{Ge} mixed systems, the average integrated absorption intensities are also higher than that of pristine \ch{RbPbI_3} in the near infrared and visible region. \\FIG S3. of the supporting information provides the variation of the real part of the dielectric function $\epsilon_{1}(\omega)$, refractive index $n(\omega)$ and reflectivity $R(\omega)$ with the variation of the photon energy. The calculated static refractive index $n(0)$ and static reflectivity $R(0)$ for all are listed in TABLE \ref{tab:e1rn}. The average static refractive indices for \ch{RbPb_{1-x}Sn_{x}I_3} systems are $2.266$, $2.287$ and $2.302$ while these values are $2.274$, $2.318$ and $2.302$ for \ch{RbPb_{1-x}Ge_{x}I_3} at $\ch{x}=0.25$, $0.50$ and $0.75$, respectively. The refractive index increases with the \ch{Sn} concentration whereas the highest static refractive index is observed for \ch{RbPb_{0.50}Ge_{0.50}I_3} in \ch{Ge} mixed systems. Similarly, the average static reflectivity increases from $15$\% to $15.5$\%  and with the increase of \ch{Sn} content while the highest $15.8$\% reflectivity is observed for all mixed systems.

\begin{table}[h!]
\caption{Effective masses for \ch{RbPb_{1-x}M_{x}I_3} when $\ch{x}=0.25$, $0.50$ and $0.25$.}
\label{tab:empsg}
\begin{tabular}{cccccccc}
\hline
\hline
\ch{M} &  & \multicolumn{2}{c}{\ch{x}=$0.25$} & \multicolumn{2}{c}{\ch{x}=$0.50$} & \multicolumn{2}{c}{\ch{x}=$0.75$} \\
\hline
& Directions & Electron & Hole & Electron & Hole & Electron & Hole \\
\hline
\multirow{ 3}{*}{\ch{Sn}} & $\Gamma \rightarrow X$ & $0.052$ & $0.225$ & $-$ & $1.294$ & $0.039$ & $0.051$ \\
& $Y \rightarrow \Gamma$ & $0.179$ & $0.620$ & $0.176$ & $0.175$ & $0.220$ & $0.088$ \\
& $\Gamma \rightarrow Z$ & $-$ & $2.662$ & $0.053$ & $0.059$ & $3.376$ & $4.635$ \\
\hline
\multirow{ 3}{*}{\ch{Ge}} & $\Gamma \rightarrow X$ & $0.155$ & $0.438$ & $-$ & $1.334$ & $0.058$ & $0.365$ \\
& $Y \rightarrow \Gamma$ & $0.154$ & $0.075$ & $0.223$ & $0.165$ &  $0.977$ & $0.232$ \\
& $\Gamma \rightarrow Z$ & $-$ & $-$ & $0.097$ & $0.073$ & $-$ & $0.370$ \\
\hline
\hline
\end{tabular}
\end{table}
 
 The charge transport property is an important factor and the effective masses ($m_{eff}$) of the charge carriers play a significant role in determining the mobility of the photogenerated electrons and holes. TABLE \ref{tab:empsg} shows the calculated effective mass for all the systems along three different crystallographic directions $\Gamma (0, 0, 0) \rightarrow X (0.5, 0, 0)$, $Y (0, 0.5, 0) \rightarrow \Gamma (0, 0, 0)$ and $\Gamma (0, 0, 0) \rightarrow Z (0, 0, 0.5)$. Effective masses $m_{eff}$ of electrons and holes are estimated by the parabolic fitting of the bottom and top of CB and VB, respectively, using the relation:
 \begin{equation}
 m_{eff}=\frac{\hbar^{2}}{\frac{\partial^{2}E(k)}{\partial k^{2}}}
\end{equation}    
where, $E(k)$ and $k$ are the band energy and the wave vector, respectively. For \ch{RbPb_{1-x}Sn_{x}I_3} systems, the average effective masses of electrons and holes are $0.116$, $1.169$; $0.115$, $0.509$; and $1.212$, $1.591$ for $x$ values of $0.25$, $0.50$ and $0.75$, respectively. On the other hand, $0.155$, $0.160$ and $0.518$ are the average effective masses of electrons while $0.257$, $0.524$ and $0.322$ are holes' average effective masses corresponding to $\ch{x}=0.25$, $0.50$ and $0.75$, respectively.   Later, the reduced effective masses ($\mu$) are calculated using $\mu=\frac{m_{e} m_{h}}{m_{e}+m_{h}}$. The estimated reduced effective masses are $0.160$, $0.094$, $0.688$ for \ch{Sn} mixed systems and $0.097$, $0.123$, $0.199$ for \ch{Ge} mixed systems corresponding to $\ch{x}=0.25$, $0.50$ and $\ch{x}=0.75$, respectively. The effective masses increase with the increase of \ch{Sn}/\ch{Ge} concentration. Effective masses also can help an exciton to dissociate into electron and hole. Hence, the binding energy ($E_{b}$) of excitons and the radius ($a^{*}$) of the lowest bound state are estimated using the relations $E_b=13.6\frac{\mu}{\epsilon_{1}(0)^{2}}$ and $a^{*}=\frac{\epsilon_{1}(0) a_{0}}{\mu}$, respectively. The estimated values of the binding energies and radii of the lowest bound state are listed in TABLE \ref{tab:e1rn}. The binding energies of \ch{Sn} and \ch{Ge} mixed systems are $54.629$, $46.702$, $333.101$ and $49.279$, $57.858$, $96.384$ meV when \ch{Sn}/\ch{Ge} concentration changes to $0.25$, $0.50$ and $0.75$, respectively. The binding energy increases with the increase of \ch{Sn}/\ch{Ge} concentrations. The radius of the lowest bound state varies from $29.444$ to $4.075$ {\AA} in \ch{RbPb_{1-x}Sn_{x}I_3} and $28.217$ to $14.086$ {\AA} in \ch{RbPb_{1-x}Ge_{x}I_3} systems. The lower binding energy and the larger radius of the lowest bound state for all the systems at $\ch{x}=0.25$ and $\ch{x}=0.50$, make the exciton to be weak and Mott-Wannier type. On the other hand, the excitons in both \ch{RbPb_{0.25}Sn_{0.75}I_3} and \ch{RbPb_{0.25}Ge_{0.75}I_3} are of Frenkel type. Therefore, excitons are found to be weak and strong for the lower and higher concentrations of \ch{Sn}/\ch{Ge}. This is consistent with the fact that the excitons in pristine \ch{RbPbI_3} are of Mott-Wannier type \cite{ref33} whereas for both the pristine \ch{RbSnI_3} and \ch{RbGeI_3} it is of Frenkel type \cite{ref41}.

\begin{figure}[h!]
\centering
  \subfloat[The SLME for \ch{RbPb_{1-x}Sn_{x}I_3}.]{\includegraphics[width=6cm,height=7cm]{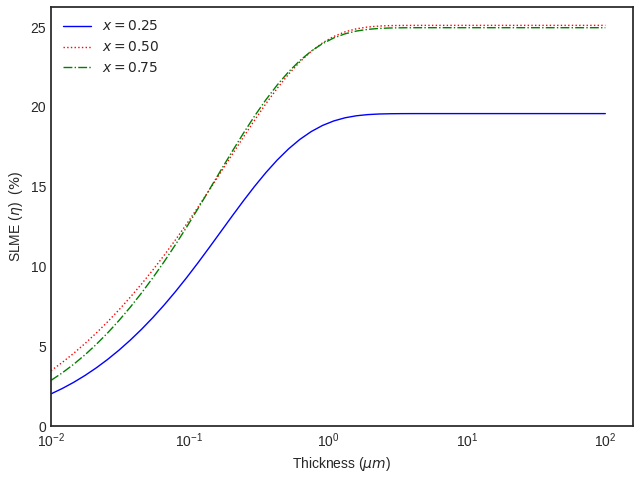}\label{fig:slmeps}}
  \hspace{0.1cm}
     \subfloat[The SLME for \ch{RbPb_{1-x}Ge_{x}I_3}.]{\includegraphics[width=6cm,height=7cm]{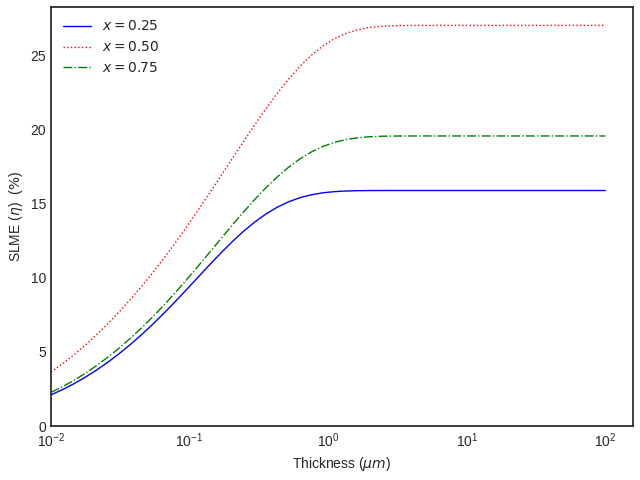}\label{fig:slmepg}}
   \caption{The SLME for both \ch{Pb}-\ch{Sn} and \ch{Pb}-\ch{Ge} mixed systems at a temperature of $300$ K}\label{fig:slmegs}
\end{figure}

Furthermore, the spectroscopic limited maximum efficiency (SLME) are estimated for all the mixed systems and it is shown in FIG \ref{fig:slmegs}. SLME is estimated as the ratio between the maximum power density ($P_m$) and the incident power density ($P_{in}$). All the required formulae to estimate the maximum power density are provided in the supporting information. The first principle estimation of the absorption coefficient, direct and indirect bandgap with PBE are used to find SLME for all the systems and these values are listed in TABLE S3 in the supporting information. The FIG. \ref{fig:slmegs} shows that SLME increases with the increase of thickness at the temperature of $300$ K. At a thickness of $500$ nm and temperature of $300$ K, SLME are estimated as $17.250$ \%, $21.897$ \% and $22.057$ \% for \ch{Pb}-\ch{Sn} mixed systems while $15.079$ \%, $23.248$ \%  and $17.358$ \% for \ch{RbPb_{1-x}Ge_{x}I_3} systems with the increase of \ch{Sn} and \ch{Ge} content, respectively. \ch{RbPb_{0.50}Ge_{0.50}I_3} exhibits the highest efficiency among all the investigated mixed systems due to the higher absorption width and the smaller bandgap.

\section{Conclusions}
In summary, the structural, electronic and optical properties of both \ch{RbPb_{1-x}Sn_{x}I_3} and \ch{RbPb_{1-x}Ge_{x}I_3} have been systematically investigated using the first principle calculations. The decrease in the volume is observed for the increase in both \ch{Sn} and \ch{Ge} content in the mixed structures except for $\ch{x}=0.50$. The highest stability is observed for \ch{RbPb_{0.50}Sn_{0.50}I_3} and \ch{RbPb_{0.50}Ge_{0.50}I_3} whereas all the systems under consideration exhibit stability. The bandstructure calculations show the indirect bandgaps for all the structures while the lowest bandgaps calculated with PBE are $1.951$ and $1.850$ eV for \ch{RbPb_{0.25}Sn_{0.75}I_3} and \ch{RbPb_{0.50}Ge_{0.50}I_3}, respectively. The decrease in the bandgaps follow an  anomalous behavior similar to the  observed in \ch{MAGe_{x}Pb_{1-x}I_3} \cite{ref28}. The comparatively highest absorption in the near infrared and visible spectra region is observed for both \ch{RbPb_{0.25}Sn_{0.75}I_3} and \ch{RbPb_{0.50}Ge_{0.50}I_3} structures. The lower effective masses of charge carriers and the binding energies of excitons are found upto $\ch{x}=0.50$ for both \ch{Pb}-\ch{Sn} and \ch{Pb}-\ch{Ge} mixed systems. The highest SLME of $22.057$\% and $23.248$\% are observed for $\ch{x}=0.75$ and $\ch{Ge}=0.50$ in case of \ch{Pb}-\ch{Sn} and \ch{Pb}-\ch{Ge} mixed systems. Among all the investigated structures, \ch{RbPb_{0.5}Ge_{0.5}I_3} exhibits the lowest bandgap, highest stability, high absorption and low exciton binding energy, thus, is the most suitable one for the photovoltaic application.

\begin{acknowledgments} 
A. N. would like to thank B. I. Sharma, Assam University, Silchar, India heartedly for the fruitful discussions and suggestions.
\end{acknowledgments}

\bibliography{refpbsnge}

\begin{thebibliography}{46}%
\makeatletter
\providecommand \@ifxundefined [1]{%
 \@ifx{#1\undefined}
}%
\providecommand \@ifnum [1]{%
 \ifnum #1\expandafter \@firstoftwo
 \else \expandafter \@secondoftwo
 \fi
}%
\providecommand \@ifx [1]{%
 \ifx #1\expandafter \@firstoftwo
 \else \expandafter \@secondoftwo
 \fi
}%
\providecommand \natexlab [1]{#1}%
\providecommand \enquote  [1]{``#1''}%
\providecommand \bibnamefont  [1]{#1}%
\providecommand \bibfnamefont [1]{#1}%
\providecommand \citenamefont [1]{#1}%
\providecommand \href@noop [0]{\@secondoftwo}%
\providecommand \href [0]{\begingroup \@sanitize@url \@href}%
\providecommand \@href[1]{\@@startlink{#1}\@@href}%
\providecommand \@@href[1]{\endgroup#1\@@endlink}%
\providecommand \@sanitize@url [0]{\catcode `\\12\catcode `\$12\catcode
  `\&12\catcode `\#12\catcode `\^12\catcode `\_12\catcode `\%12\relax}%
\providecommand \@@startlink[1]{}%
\providecommand \@@endlink[0]{}%
\providecommand \url  [0]{\begingroup\@sanitize@url \@url }%
\providecommand \@url [1]{\endgroup\@href {#1}{\urlprefix }}%
\providecommand \urlprefix  [0]{URL }%
\providecommand \Eprint [0]{\href }%
\providecommand \doibase [0]{https://doi.org/}%
\providecommand \selectlanguage [0]{\@gobble}%
\providecommand \bibinfo  [0]{\@secondoftwo}%
\providecommand \bibfield  [0]{\@secondoftwo}%
\providecommand \translation [1]{[#1]}%
\providecommand \BibitemOpen [0]{}%
\providecommand \bibitemStop [0]{}%
\providecommand \bibitemNoStop [0]{.\EOS\space}%
\providecommand \EOS [0]{\spacefactor3000\relax}%
\providecommand \BibitemShut  [1]{\csname bibitem#1\endcsname}%
\let\auto@bib@innerbib\@empty
\bibitem [{\citenamefont {Baikie}\ \emph {et~al.}(2013)\citenamefont {Baikie},
  \citenamefont {Fang}, \citenamefont {Kadro}, \citenamefont {Schreyer},
  \citenamefont {Wei}, \citenamefont {Mhaisalkar}, \citenamefont {Graetzel},\
  and\ \citenamefont {White}}]{ref4}%
  \BibitemOpen
  \bibfield  {author} {\bibinfo {author} {\bibfnamefont {T.}~\bibnamefont
  {Baikie}}, \bibinfo {author} {\bibfnamefont {Y.}~\bibnamefont {Fang}},
  \bibinfo {author} {\bibfnamefont {J.~M.}\ \bibnamefont {Kadro}}, \bibinfo
  {author} {\bibfnamefont {M.}~\bibnamefont {Schreyer}}, \bibinfo {author}
  {\bibfnamefont {F.}~\bibnamefont {Wei}}, \bibinfo {author} {\bibfnamefont
  {S.~G.}\ \bibnamefont {Mhaisalkar}}, \bibinfo {author} {\bibfnamefont
  {M.}~\bibnamefont {Graetzel}},\ and\ \bibinfo {author} {\bibfnamefont
  {T.~J.}\ \bibnamefont {White}},\ }\bibfield  {title} {\bibinfo {title}
  {Synthesis and crystal chemistry of the hybrid perovskite \ch{CH_3NH_3PbI_3}
  for solid-state sensitised solar cell applications},\ }\href@noop {}
  {\bibfield  {journal} {\bibinfo  {journal} {Journal of Materials Chemistry
  A}\ }\textbf {\bibinfo {volume} {1}},\ \bibinfo {pages} {5628} (\bibinfo
  {year} {2013})}\BibitemShut {NoStop}%
\bibitem [{\citenamefont {Shao}\ \emph {et~al.}(2014)\citenamefont {Shao},
  \citenamefont {Xiao}, \citenamefont {Bi}, \citenamefont {Yuan},\ and\
  \citenamefont {Huang}}]{ref5}%
  \BibitemOpen
  \bibfield  {author} {\bibinfo {author} {\bibfnamefont {Y.}~\bibnamefont
  {Shao}}, \bibinfo {author} {\bibfnamefont {Z.}~\bibnamefont {Xiao}}, \bibinfo
  {author} {\bibfnamefont {C.}~\bibnamefont {Bi}}, \bibinfo {author}
  {\bibfnamefont {Y.}~\bibnamefont {Yuan}},\ and\ \bibinfo {author}
  {\bibfnamefont {J.}~\bibnamefont {Huang}},\ }\bibfield  {title} {\bibinfo
  {title} {Origin and elimination of photocurrent hysteresis by fullerene
  passivation in \ch{CH_3NH_3PbI_3} planar heterojunction solar cells},\
  }\href@noop {} {\bibfield  {journal} {\bibinfo  {journal} {Nature
  communications}\ }\textbf {\bibinfo {volume} {5}},\ \bibinfo {pages} {1}
  (\bibinfo {year} {2014})}\BibitemShut {NoStop}%
\bibitem [{\citenamefont {D’innocenzo}\ \emph {et~al.}(2014)\citenamefont
  {D’innocenzo}, \citenamefont {Grancini}, \citenamefont {Alcocer},
  \citenamefont {Kandada}, \citenamefont {Stranks}, \citenamefont {Lee},
  \citenamefont {Lanzani}, \citenamefont {Snaith},\ and\ \citenamefont
  {Petrozza}}]{ref6}%
  \BibitemOpen
  \bibfield  {author} {\bibinfo {author} {\bibfnamefont {V.}~\bibnamefont
  {D’innocenzo}}, \bibinfo {author} {\bibfnamefont {G.}~\bibnamefont
  {Grancini}}, \bibinfo {author} {\bibfnamefont {M.~J.}\ \bibnamefont
  {Alcocer}}, \bibinfo {author} {\bibfnamefont {A.~R.~S.}\ \bibnamefont
  {Kandada}}, \bibinfo {author} {\bibfnamefont {S.~D.}\ \bibnamefont
  {Stranks}}, \bibinfo {author} {\bibfnamefont {M.~M.}\ \bibnamefont {Lee}},
  \bibinfo {author} {\bibfnamefont {G.}~\bibnamefont {Lanzani}}, \bibinfo
  {author} {\bibfnamefont {H.~J.}\ \bibnamefont {Snaith}},\ and\ \bibinfo
  {author} {\bibfnamefont {A.}~\bibnamefont {Petrozza}},\ }\bibfield  {title}
  {\bibinfo {title} {Excitons versus free charges in organo-lead tri-halide
  perovskites},\ }\href@noop {} {\bibfield  {journal} {\bibinfo  {journal}
  {Nature communications}\ }\textbf {\bibinfo {volume} {5}},\ \bibinfo {pages}
  {1} (\bibinfo {year} {2014})}\BibitemShut {NoStop}%
\bibitem [{\citenamefont {Yang}\ \emph
  {et~al.}(2017{\natexlab{a}})\citenamefont {Yang}, \citenamefont {Park},
  \citenamefont {Jung}, \citenamefont {Jeon}, \citenamefont {Kim},
  \citenamefont {Lee}, \citenamefont {Shin}, \citenamefont {Seo}, \citenamefont
  {Kim}, \citenamefont {Noh} \emph {et~al.}}]{ref7}%
  \BibitemOpen
  \bibfield  {author} {\bibinfo {author} {\bibfnamefont {W.~S.}\ \bibnamefont
  {Yang}}, \bibinfo {author} {\bibfnamefont {B.-W.}\ \bibnamefont {Park}},
  \bibinfo {author} {\bibfnamefont {E.~H.}\ \bibnamefont {Jung}}, \bibinfo
  {author} {\bibfnamefont {N.~J.}\ \bibnamefont {Jeon}}, \bibinfo {author}
  {\bibfnamefont {Y.~C.}\ \bibnamefont {Kim}}, \bibinfo {author} {\bibfnamefont
  {D.~U.}\ \bibnamefont {Lee}}, \bibinfo {author} {\bibfnamefont {S.~S.}\
  \bibnamefont {Shin}}, \bibinfo {author} {\bibfnamefont {J.}~\bibnamefont
  {Seo}}, \bibinfo {author} {\bibfnamefont {E.~K.}\ \bibnamefont {Kim}},
  \bibinfo {author} {\bibfnamefont {J.~H.}\ \bibnamefont {Noh}}, \emph
  {et~al.},\ }\bibfield  {title} {\bibinfo {title} {Iodide management in
  formamidinium-lead-halide--based perovskite layers for efficient solar
  cells},\ }\href@noop {} {\bibfield  {journal} {\bibinfo  {journal} {Science}\
  }\textbf {\bibinfo {volume} {356}},\ \bibinfo {pages} {1376} (\bibinfo {year}
  {2017}{\natexlab{a}})}\BibitemShut {NoStop}%
\bibitem [{\citenamefont {Seo}\ \emph {et~al.}(2016)\citenamefont {Seo},
  \citenamefont {Noh},\ and\ \citenamefont {Seok}}]{ref8}%
  \BibitemOpen
  \bibfield  {author} {\bibinfo {author} {\bibfnamefont {J.}~\bibnamefont
  {Seo}}, \bibinfo {author} {\bibfnamefont {J.~H.}\ \bibnamefont {Noh}},\ and\
  \bibinfo {author} {\bibfnamefont {S.~I.}\ \bibnamefont {Seok}},\ }\bibfield
  {title} {\bibinfo {title} {Rational strategies for efficient perovskite solar
  cells},\ }\href@noop {} {\bibfield  {journal} {\bibinfo  {journal} {Accounts
  of chemical research}\ }\textbf {\bibinfo {volume} {49}},\ \bibinfo {pages}
  {562} (\bibinfo {year} {2016})}\BibitemShut {NoStop}%
\bibitem [{\citenamefont {Dunlap-Shohl}\ \emph {et~al.}(2018)\citenamefont
  {Dunlap-Shohl}, \citenamefont {Zhou}, \citenamefont {Padture},\ and\
  \citenamefont {Mitzi}}]{ref9}%
  \BibitemOpen
  \bibfield  {author} {\bibinfo {author} {\bibfnamefont {W.~A.}\ \bibnamefont
  {Dunlap-Shohl}}, \bibinfo {author} {\bibfnamefont {Y.}~\bibnamefont {Zhou}},
  \bibinfo {author} {\bibfnamefont {N.~P.}\ \bibnamefont {Padture}},\ and\
  \bibinfo {author} {\bibfnamefont {D.~B.}\ \bibnamefont {Mitzi}},\ }\bibfield
  {title} {\bibinfo {title} {Synthetic approaches for halide perovskite thin
  films},\ }\href@noop {} {\bibfield  {journal} {\bibinfo  {journal} {Chemical
  reviews}\ }\textbf {\bibinfo {volume} {119}},\ \bibinfo {pages} {3193}
  (\bibinfo {year} {2018})}\BibitemShut {NoStop}%
\bibitem [{\citenamefont {Meng}\ \emph {et~al.}(2018)\citenamefont {Meng},
  \citenamefont {Cui}, \citenamefont {Rager}, \citenamefont {Zhang},
  \citenamefont {Wang}, \citenamefont {Yu}, \citenamefont {Harn}, \citenamefont
  {Kang}, \citenamefont {Wagner}, \citenamefont {Liu} \emph {et~al.}}]{ref10}%
  \BibitemOpen
  \bibfield  {author} {\bibinfo {author} {\bibfnamefont {X.}~\bibnamefont
  {Meng}}, \bibinfo {author} {\bibfnamefont {X.}~\bibnamefont {Cui}}, \bibinfo
  {author} {\bibfnamefont {M.}~\bibnamefont {Rager}}, \bibinfo {author}
  {\bibfnamefont {S.}~\bibnamefont {Zhang}}, \bibinfo {author} {\bibfnamefont
  {Z.}~\bibnamefont {Wang}}, \bibinfo {author} {\bibfnamefont {J.}~\bibnamefont
  {Yu}}, \bibinfo {author} {\bibfnamefont {Y.~W.}\ \bibnamefont {Harn}},
  \bibinfo {author} {\bibfnamefont {Z.}~\bibnamefont {Kang}}, \bibinfo {author}
  {\bibfnamefont {B.~K.}\ \bibnamefont {Wagner}}, \bibinfo {author}
  {\bibfnamefont {Y.}~\bibnamefont {Liu}}, \emph {et~al.},\ }\bibfield  {title}
  {\bibinfo {title} {Cascade charge transfer enabled by incorporating
  edge-enriched graphene nanoribbons for mesostructured perovskite solar cells
  with enhanced performance},\ }\href@noop {} {\bibfield  {journal} {\bibinfo
  {journal} {Nano Energy}\ }\textbf {\bibinfo {volume} {52}},\ \bibinfo {pages}
  {123} (\bibinfo {year} {2018})}\BibitemShut {NoStop}%
\bibitem [{\citenamefont {Dong}\ \emph {et~al.}(2019)\citenamefont {Dong},
  \citenamefont {Xi}, \citenamefont {Zuo}, \citenamefont {Li}, \citenamefont
  {Yang}, \citenamefont {Wang}, \citenamefont {Yu}, \citenamefont {Ma},
  \citenamefont {Ran}, \citenamefont {Gao} \emph {et~al.}}]{ref11}%
  \BibitemOpen
  \bibfield  {author} {\bibinfo {author} {\bibfnamefont {H.}~\bibnamefont
  {Dong}}, \bibinfo {author} {\bibfnamefont {J.}~\bibnamefont {Xi}}, \bibinfo
  {author} {\bibfnamefont {L.}~\bibnamefont {Zuo}}, \bibinfo {author}
  {\bibfnamefont {J.}~\bibnamefont {Li}}, \bibinfo {author} {\bibfnamefont
  {Y.}~\bibnamefont {Yang}}, \bibinfo {author} {\bibfnamefont {D.}~\bibnamefont
  {Wang}}, \bibinfo {author} {\bibfnamefont {Y.}~\bibnamefont {Yu}}, \bibinfo
  {author} {\bibfnamefont {L.}~\bibnamefont {Ma}}, \bibinfo {author}
  {\bibfnamefont {C.}~\bibnamefont {Ran}}, \bibinfo {author} {\bibfnamefont
  {W.}~\bibnamefont {Gao}}, \emph {et~al.},\ }\bibfield  {title} {\bibinfo
  {title} {Conjugated molecules “bridge”: Functional ligand toward highly
  efficient and long-term stable perovskite solar cell},\ }\href@noop {}
  {\bibfield  {journal} {\bibinfo  {journal} {Advanced Functional Materials}\
  }\textbf {\bibinfo {volume} {29}},\ \bibinfo {pages} {1808119} (\bibinfo
  {year} {2019})}\BibitemShut {NoStop}%
\bibitem [{\citenamefont {Arora}\ \emph {et~al.}(2017)\citenamefont {Arora},
  \citenamefont {Dar}, \citenamefont {Hinderhofer}, \citenamefont {Pellet},
  \citenamefont {Schreiber}, \citenamefont {Zakeeruddin},\ and\ \citenamefont
  {Gr{\"a}tzel}}]{ref12}%
  \BibitemOpen
  \bibfield  {author} {\bibinfo {author} {\bibfnamefont {N.}~\bibnamefont
  {Arora}}, \bibinfo {author} {\bibfnamefont {M.~I.}\ \bibnamefont {Dar}},
  \bibinfo {author} {\bibfnamefont {A.}~\bibnamefont {Hinderhofer}}, \bibinfo
  {author} {\bibfnamefont {N.}~\bibnamefont {Pellet}}, \bibinfo {author}
  {\bibfnamefont {F.}~\bibnamefont {Schreiber}}, \bibinfo {author}
  {\bibfnamefont {S.~M.}\ \bibnamefont {Zakeeruddin}},\ and\ \bibinfo {author}
  {\bibfnamefont {M.}~\bibnamefont {Gr{\"a}tzel}},\ }\bibfield  {title}
  {\bibinfo {title} {Perovskite solar cells with cuscn hole extraction layers
  yield stabilized efficiencies greater than 20\%},\ }\href@noop {} {\bibfield
  {journal} {\bibinfo  {journal} {Science}\ }\textbf {\bibinfo {volume}
  {358}},\ \bibinfo {pages} {768} (\bibinfo {year} {2017})}\BibitemShut
  {NoStop}%
\bibitem [{\citenamefont {Fu}\ \emph {et~al.}(2017)\citenamefont {Fu},
  \citenamefont {Bao}, \citenamefont {Zhang}, \citenamefont {Zhang},
  \citenamefont {Ke}, \citenamefont {Lin}, \citenamefont {Dai}, \citenamefont
  {Huang},\ and\ \citenamefont {Lei}}]{ref13}%
  \BibitemOpen
  \bibfield  {author} {\bibinfo {author} {\bibfnamefont {N.}~\bibnamefont
  {Fu}}, \bibinfo {author} {\bibfnamefont {Z.~Y.}\ \bibnamefont {Bao}},
  \bibinfo {author} {\bibfnamefont {Y.-L.}\ \bibnamefont {Zhang}}, \bibinfo
  {author} {\bibfnamefont {G.}~\bibnamefont {Zhang}}, \bibinfo {author}
  {\bibfnamefont {S.}~\bibnamefont {Ke}}, \bibinfo {author} {\bibfnamefont
  {P.}~\bibnamefont {Lin}}, \bibinfo {author} {\bibfnamefont {J.}~\bibnamefont
  {Dai}}, \bibinfo {author} {\bibfnamefont {H.}~\bibnamefont {Huang}},\ and\
  \bibinfo {author} {\bibfnamefont {D.~Y.}\ \bibnamefont {Lei}},\ }\bibfield
  {title} {\bibinfo {title} {Panchromatic thin perovskite solar cells with
  broadband plasmonic absorption enhancement and efficient light scattering
  management by \ch{Au}@\ch{Ag} core-shell nanocuboids},\ }\href@noop {}
  {\bibfield  {journal} {\bibinfo  {journal} {Nano Energy}\ }\textbf {\bibinfo
  {volume} {41}},\ \bibinfo {pages} {654} (\bibinfo {year} {2017})}\BibitemShut
  {NoStop}%
\bibitem [{\citenamefont {Buin}\ \emph {et~al.}(2015)\citenamefont {Buin},
  \citenamefont {Comin}, \citenamefont {Xu}, \citenamefont {Ip},\ and\
  \citenamefont {Sargent}}]{ref14}%
  \BibitemOpen
  \bibfield  {author} {\bibinfo {author} {\bibfnamefont {A.}~\bibnamefont
  {Buin}}, \bibinfo {author} {\bibfnamefont {R.}~\bibnamefont {Comin}},
  \bibinfo {author} {\bibfnamefont {J.}~\bibnamefont {Xu}}, \bibinfo {author}
  {\bibfnamefont {A.~H.}\ \bibnamefont {Ip}},\ and\ \bibinfo {author}
  {\bibfnamefont {E.~H.}\ \bibnamefont {Sargent}},\ }\bibfield  {title}
  {\bibinfo {title} {Halide-dependent electronic structure of organolead
  perovskite materials},\ }\href@noop {} {\bibfield  {journal} {\bibinfo
  {journal} {Chemistry of Materials}\ }\textbf {\bibinfo {volume} {27}},\
  \bibinfo {pages} {4405} (\bibinfo {year} {2015})}\BibitemShut {NoStop}%
\bibitem [{\citenamefont {Misra}\ \emph {et~al.}(2015)\citenamefont {Misra},
  \citenamefont {Aharon}, \citenamefont {Li}, \citenamefont {Mogilyansky},
  \citenamefont {Visoly-Fisher}, \citenamefont {Etgar},\ and\ \citenamefont
  {Katz}}]{ref15}%
  \BibitemOpen
  \bibfield  {author} {\bibinfo {author} {\bibfnamefont {R.~K.}\ \bibnamefont
  {Misra}}, \bibinfo {author} {\bibfnamefont {S.}~\bibnamefont {Aharon}},
  \bibinfo {author} {\bibfnamefont {B.}~\bibnamefont {Li}}, \bibinfo {author}
  {\bibfnamefont {D.}~\bibnamefont {Mogilyansky}}, \bibinfo {author}
  {\bibfnamefont {I.}~\bibnamefont {Visoly-Fisher}}, \bibinfo {author}
  {\bibfnamefont {L.}~\bibnamefont {Etgar}},\ and\ \bibinfo {author}
  {\bibfnamefont {E.~A.}\ \bibnamefont {Katz}},\ }\bibfield  {title} {\bibinfo
  {title} {Temperature-and component-dependent degradation of perovskite
  photovoltaic materials under concentrated sunlight},\ }\href@noop {}
  {\bibfield  {journal} {\bibinfo  {journal} {The journal of physical chemistry
  letters}\ }\textbf {\bibinfo {volume} {6}},\ \bibinfo {pages} {326} (\bibinfo
  {year} {2015})}\BibitemShut {NoStop}%
\bibitem [{\citenamefont {Xiang}\ and\ \citenamefont {Tress}(2019)}]{ref16}%
  \BibitemOpen
  \bibfield  {author} {\bibinfo {author} {\bibfnamefont {W.}~\bibnamefont
  {Xiang}}\ and\ \bibinfo {author} {\bibfnamefont {W.}~\bibnamefont {Tress}},\
  }\bibfield  {title} {\bibinfo {title} {Review on recent progress of
  all-inorganic metal halide perovskites and solar cells},\ }\href@noop {}
  {\bibfield  {journal} {\bibinfo  {journal} {Advanced Materials}\ }\textbf
  {\bibinfo {volume} {31}},\ \bibinfo {pages} {1902851} (\bibinfo {year}
  {2019})}\BibitemShut {NoStop}%
\bibitem [{\citenamefont {Dang}\ \emph {et~al.}(2016)\citenamefont {Dang},
  \citenamefont {Ju}, \citenamefont {Wang},\ and\ \citenamefont {Tao}}]{ref17}%
  \BibitemOpen
  \bibfield  {author} {\bibinfo {author} {\bibfnamefont {Y.}~\bibnamefont
  {Dang}}, \bibinfo {author} {\bibfnamefont {D.}~\bibnamefont {Ju}}, \bibinfo
  {author} {\bibfnamefont {L.}~\bibnamefont {Wang}},\ and\ \bibinfo {author}
  {\bibfnamefont {X.}~\bibnamefont {Tao}},\ }\bibfield  {title} {\bibinfo
  {title} {Recent progress in the synthesis of hybrid halide perovskite single
  crystals},\ }\href@noop {} {\bibfield  {journal} {\bibinfo  {journal}
  {CrystEngComm}\ }\textbf {\bibinfo {volume} {18}},\ \bibinfo {pages} {4476}
  (\bibinfo {year} {2016})}\BibitemShut {NoStop}%
\bibitem [{\citenamefont {Wang}\ \emph {et~al.}(2014)\citenamefont {Wang},
  \citenamefont {Xiao},\ and\ \citenamefont {Chen}}]{ref18}%
  \BibitemOpen
  \bibfield  {author} {\bibinfo {author} {\bibfnamefont {B.}~\bibnamefont
  {Wang}}, \bibinfo {author} {\bibfnamefont {X.}~\bibnamefont {Xiao}},\ and\
  \bibinfo {author} {\bibfnamefont {T.}~\bibnamefont {Chen}},\ }\bibfield
  {title} {\bibinfo {title} {Perovskite photovoltaics: a high-efficiency
  newcomer to the solar cell family},\ }\href@noop {} {\bibfield  {journal}
  {\bibinfo  {journal} {Nanoscale}\ }\textbf {\bibinfo {volume} {6}},\ \bibinfo
  {pages} {12287} (\bibinfo {year} {2014})}\BibitemShut {NoStop}%
\bibitem [{\citenamefont {Yang}\ \emph
  {et~al.}(2017{\natexlab{b}})\citenamefont {Yang}, \citenamefont {Skelton},
  \citenamefont {Da~Silva}, \citenamefont {Frost},\ and\ \citenamefont
  {Walsh}}]{ref19}%
  \BibitemOpen
  \bibfield  {author} {\bibinfo {author} {\bibfnamefont {R.~X.}\ \bibnamefont
  {Yang}}, \bibinfo {author} {\bibfnamefont {J.~M.}\ \bibnamefont {Skelton}},
  \bibinfo {author} {\bibfnamefont {E.~L.}\ \bibnamefont {Da~Silva}}, \bibinfo
  {author} {\bibfnamefont {J.~M.}\ \bibnamefont {Frost}},\ and\ \bibinfo
  {author} {\bibfnamefont {A.}~\bibnamefont {Walsh}},\ }\bibfield  {title}
  {\bibinfo {title} {Spontaneous octahedral tilting in the cubic inorganic
  cesium halide perovskites \ch{CsSnX_3} and \ch{CsPbX_3} (\ch{X}= \ch{F},
  \ch{Cl}, \ch{Br}, \ch{I})},\ }\href@noop {} {\bibfield  {journal} {\bibinfo
  {journal} {The journal of physical chemistry letters}\ }\textbf {\bibinfo
  {volume} {8}},\ \bibinfo {pages} {4720} (\bibinfo {year}
  {2017}{\natexlab{b}})}\BibitemShut {NoStop}%
\bibitem [{\citenamefont {Hoffman}\ \emph {et~al.}(2016)\citenamefont
  {Hoffman}, \citenamefont {Schleper},\ and\ \citenamefont {Kamat}}]{ref20}%
  \BibitemOpen
  \bibfield  {author} {\bibinfo {author} {\bibfnamefont {J.~B.}\ \bibnamefont
  {Hoffman}}, \bibinfo {author} {\bibfnamefont {A.~L.}\ \bibnamefont
  {Schleper}},\ and\ \bibinfo {author} {\bibfnamefont {P.~V.}\ \bibnamefont
  {Kamat}},\ }\bibfield  {title} {\bibinfo {title} {Transformation of sintered
  cspbbr3 nanocrystals to cubic \ch{CsPbI_3} and gradient \ch{CsPbBr_xI_{3-x}}
  through halide exchange},\ }\href@noop {} {\bibfield  {journal} {\bibinfo
  {journal} {Journal of the American Chemical Society}\ }\textbf {\bibinfo
  {volume} {138}},\ \bibinfo {pages} {8603} (\bibinfo {year}
  {2016})}\BibitemShut {NoStop}%
\bibitem [{\citenamefont {Kulbak}\ \emph {et~al.}(2015)\citenamefont {Kulbak},
  \citenamefont {Cahen},\ and\ \citenamefont {Hodes}}]{ref21}%
  \BibitemOpen
  \bibfield  {author} {\bibinfo {author} {\bibfnamefont {M.}~\bibnamefont
  {Kulbak}}, \bibinfo {author} {\bibfnamefont {D.}~\bibnamefont {Cahen}},\ and\
  \bibinfo {author} {\bibfnamefont {G.}~\bibnamefont {Hodes}},\ }\bibfield
  {title} {\bibinfo {title} {How important is the organic part of lead halide
  perovskite photovoltaic cells? efficient \ch{CsPbBr_3} cells},\ }\href@noop
  {} {\bibfield  {journal} {\bibinfo  {journal} {The journal of physical
  chemistry letters}\ }\textbf {\bibinfo {volume} {6}},\ \bibinfo {pages}
  {2452} (\bibinfo {year} {2015})}\BibitemShut {NoStop}%
\bibitem [{\citenamefont {Kovalsky}\ \emph {et~al.}(2017)\citenamefont
  {Kovalsky}, \citenamefont {Wang}, \citenamefont {Marek}, \citenamefont
  {Burda},\ and\ \citenamefont {Dyck}}]{ref22}%
  \BibitemOpen
  \bibfield  {author} {\bibinfo {author} {\bibfnamefont {A.}~\bibnamefont
  {Kovalsky}}, \bibinfo {author} {\bibfnamefont {L.}~\bibnamefont {Wang}},
  \bibinfo {author} {\bibfnamefont {G.~T.}\ \bibnamefont {Marek}}, \bibinfo
  {author} {\bibfnamefont {C.}~\bibnamefont {Burda}},\ and\ \bibinfo {author}
  {\bibfnamefont {J.~S.}\ \bibnamefont {Dyck}},\ }\bibfield  {title} {\bibinfo
  {title} {Thermal conductivity of \ch{CH_3NH_3PbI_3} and \ch{CsPbI_3}:
  Measuring the effect of the methylammonium ion on phonon scattering},\
  }\href@noop {} {\bibfield  {journal} {\bibinfo  {journal} {The Journal of
  Physical Chemistry C}\ }\textbf {\bibinfo {volume} {121}},\ \bibinfo {pages}
  {3228} (\bibinfo {year} {2017})}\BibitemShut {NoStop}%
\bibitem [{\citenamefont {Chung}\ \emph {et~al.}(2012)\citenamefont {Chung},
  \citenamefont {Song}, \citenamefont {Im}, \citenamefont {Androulakis},
  \citenamefont {Malliakas}, \citenamefont {Li}, \citenamefont {Freeman},
  \citenamefont {Kenney},\ and\ \citenamefont {Kanatzidis}}]{ref23}%
  \BibitemOpen
  \bibfield  {author} {\bibinfo {author} {\bibfnamefont {I.}~\bibnamefont
  {Chung}}, \bibinfo {author} {\bibfnamefont {J.-H.}\ \bibnamefont {Song}},
  \bibinfo {author} {\bibfnamefont {J.}~\bibnamefont {Im}}, \bibinfo {author}
  {\bibfnamefont {J.}~\bibnamefont {Androulakis}}, \bibinfo {author}
  {\bibfnamefont {C.~D.}\ \bibnamefont {Malliakas}}, \bibinfo {author}
  {\bibfnamefont {H.}~\bibnamefont {Li}}, \bibinfo {author} {\bibfnamefont
  {A.~J.}\ \bibnamefont {Freeman}}, \bibinfo {author} {\bibfnamefont {J.~T.}\
  \bibnamefont {Kenney}},\ and\ \bibinfo {author} {\bibfnamefont {M.~G.}\
  \bibnamefont {Kanatzidis}},\ }\bibfield  {title} {\bibinfo {title}
  {\ch{CsSnI_3}: semiconductor or metal? high electrical conductivity and
  strong near-infrared photoluminescence from a single material. high hole
  mobility and phase-transitions},\ }\href@noop {} {\bibfield  {journal}
  {\bibinfo  {journal} {Journal of the American Chemical Society}\ }\textbf
  {\bibinfo {volume} {134}},\ \bibinfo {pages} {8579} (\bibinfo {year}
  {2012})}\BibitemShut {NoStop}%
\bibitem [{\citenamefont {Brgoch}\ \emph {et~al.}(2014)\citenamefont {Brgoch},
  \citenamefont {Lehner}, \citenamefont {Chabinyc},\ and\ \citenamefont
  {Seshadri}}]{ref46}%
  \BibitemOpen
  \bibfield  {author} {\bibinfo {author} {\bibfnamefont {J.}~\bibnamefont
  {Brgoch}}, \bibinfo {author} {\bibfnamefont {A.~J.}\ \bibnamefont {Lehner}},
  \bibinfo {author} {\bibfnamefont {M.}~\bibnamefont {Chabinyc}},\ and\
  \bibinfo {author} {\bibfnamefont {R.}~\bibnamefont {Seshadri}},\ }\bibfield
  {title} {\bibinfo {title} {Ab initio calculations of band gaps and absolute
  band positions of polymorphs of \ch{RbPbI_3} and \ch{CsPbI_3}: implications
  for main-group halide perovskite photovoltaics},\ }\href@noop {} {\bibfield
  {journal} {\bibinfo  {journal} {The Journal of Physical Chemistry C}\
  }\textbf {\bibinfo {volume} {118}},\ \bibinfo {pages} {27721} (\bibinfo
  {year} {2014})}\BibitemShut {NoStop}%
\bibitem [{\citenamefont {Liu}\ \emph {et~al.}(2019)\citenamefont {Liu},
  \citenamefont {Sa},\ and\ \citenamefont {Wu}}]{ref47}%
  \BibitemOpen
  \bibfield  {author} {\bibinfo {author} {\bibfnamefont {D.}~\bibnamefont
  {Liu}}, \bibinfo {author} {\bibfnamefont {R.}~\bibnamefont {Sa}},\ and\
  \bibinfo {author} {\bibfnamefont {K.}~\bibnamefont {Wu}},\ }\bibfield
  {title} {\bibinfo {title} {First-principles insight on the electronic and
  optical properties of \ch{Ge}-based inorganic perovskites},\ }\href@noop {}
  {\bibfield  {journal} {\bibinfo  {journal} {Applied Physics Express}\
  }\textbf {\bibinfo {volume} {12}},\ \bibinfo {pages} {071007} (\bibinfo
  {year} {2019})}\BibitemShut {NoStop}%
\bibitem [{\citenamefont {Nyayban}\ \emph {et~al.}(2020)\citenamefont
  {Nyayban}, \citenamefont {Panda}, \citenamefont {Chowdhury},\ and\
  \citenamefont {Sharma}}]{ref33}%
  \BibitemOpen
  \bibfield  {author} {\bibinfo {author} {\bibfnamefont {A.}~\bibnamefont
  {Nyayban}}, \bibinfo {author} {\bibfnamefont {S.}~\bibnamefont {Panda}},
  \bibinfo {author} {\bibfnamefont {A.}~\bibnamefont {Chowdhury}},\ and\
  \bibinfo {author} {\bibfnamefont {B.~I.}\ \bibnamefont {Sharma}},\ }\bibfield
   {title} {\bibinfo {title} {First principle studies of rubidium lead halides
  towards photovoltaic application},\ }\href@noop {} {\bibfield  {journal}
  {\bibinfo  {journal} {Materials Today Communications}\ }\textbf {\bibinfo
  {volume} {24}},\ \bibinfo {pages} {101190} (\bibinfo {year}
  {2020})}\BibitemShut {NoStop}%
\bibitem [{\citenamefont {Nyayban}\ \emph {et~al.}(2021)\citenamefont
  {Nyayban}, \citenamefont {Panda},\ and\ \citenamefont {Chowdhury}}]{ref41}%
  \BibitemOpen
  \bibfield  {author} {\bibinfo {author} {\bibfnamefont {A.}~\bibnamefont
  {Nyayban}}, \bibinfo {author} {\bibfnamefont {S.}~\bibnamefont {Panda}},\
  and\ \bibinfo {author} {\bibfnamefont {A.}~\bibnamefont {Chowdhury}},\
  }\bibfield  {title} {\bibinfo {title} {Structural, electronic and optical
  properties of lead free rb based triiodide for photovoltaic application: an
  ab initio study},\ }\href@noop {} {\bibfield  {journal} {\bibinfo  {journal}
  {Journal of Physics: Condensed Matter}\ }\textbf {\bibinfo {volume} {33}},\
  \bibinfo {pages} {375702} (\bibinfo {year} {2021})}\BibitemShut {NoStop}%
\bibitem [{\citenamefont {Liu}\ \emph {et~al.}(2016)\citenamefont {Liu},
  \citenamefont {Yang}, \citenamefont {Chueh}, \citenamefont {Rajagopal},
  \citenamefont {Williams}, \citenamefont {Sun},\ and\ \citenamefont
  {Jen}}]{ref24}%
  \BibitemOpen
  \bibfield  {author} {\bibinfo {author} {\bibfnamefont {X.}~\bibnamefont
  {Liu}}, \bibinfo {author} {\bibfnamefont {Z.}~\bibnamefont {Yang}}, \bibinfo
  {author} {\bibfnamefont {C.-C.}\ \bibnamefont {Chueh}}, \bibinfo {author}
  {\bibfnamefont {A.}~\bibnamefont {Rajagopal}}, \bibinfo {author}
  {\bibfnamefont {S.~T.}\ \bibnamefont {Williams}}, \bibinfo {author}
  {\bibfnamefont {Y.}~\bibnamefont {Sun}},\ and\ \bibinfo {author}
  {\bibfnamefont {A.~K.-Y.}\ \bibnamefont {Jen}},\ }\bibfield  {title}
  {\bibinfo {title} {Improved efficiency and stability of \ch{Pb}-\ch{Sn}
  binary perovskite solar cells by \ch{Cs} substitution},\ }\href@noop {}
  {\bibfield  {journal} {\bibinfo  {journal} {Journal of Materials Chemistry
  A}\ }\textbf {\bibinfo {volume} {4}},\ \bibinfo {pages} {17939} (\bibinfo
  {year} {2016})}\BibitemShut {NoStop}%
\bibitem [{\citenamefont {Leijtens}\ \emph {et~al.}(2017)\citenamefont
  {Leijtens}, \citenamefont {Prasanna}, \citenamefont {Gold-Parker},
  \citenamefont {Toney},\ and\ \citenamefont {McGehee}}]{ref25}%
  \BibitemOpen
  \bibfield  {author} {\bibinfo {author} {\bibfnamefont {T.}~\bibnamefont
  {Leijtens}}, \bibinfo {author} {\bibfnamefont {R.}~\bibnamefont {Prasanna}},
  \bibinfo {author} {\bibfnamefont {A.}~\bibnamefont {Gold-Parker}}, \bibinfo
  {author} {\bibfnamefont {M.~F.}\ \bibnamefont {Toney}},\ and\ \bibinfo
  {author} {\bibfnamefont {M.~D.}\ \bibnamefont {McGehee}},\ }\bibfield
  {title} {\bibinfo {title} {Mechanism of tin oxidation and stabilization by
  lead substitution in tin halide perovskites},\ }\href@noop {} {\bibfield
  {journal} {\bibinfo  {journal} {ACS Energy Letters}\ }\textbf {\bibinfo
  {volume} {2}},\ \bibinfo {pages} {2159} (\bibinfo {year} {2017})}\BibitemShut
  {NoStop}%
\bibitem [{\citenamefont {Hao}\ \emph {et~al.}(2014{\natexlab{a}})\citenamefont
  {Hao}, \citenamefont {Stoumpos}, \citenamefont {Chang},\ and\ \citenamefont
  {Kanatzidis}}]{ref26}%
  \BibitemOpen
  \bibfield  {author} {\bibinfo {author} {\bibfnamefont {F.}~\bibnamefont
  {Hao}}, \bibinfo {author} {\bibfnamefont {C.~C.}\ \bibnamefont {Stoumpos}},
  \bibinfo {author} {\bibfnamefont {R.~P.}\ \bibnamefont {Chang}},\ and\
  \bibinfo {author} {\bibfnamefont {M.~G.}\ \bibnamefont {Kanatzidis}},\
  }\bibfield  {title} {\bibinfo {title} {Anomalous band gap behavior in mixed
  \ch{Sn} and \ch{Pb} perovskites enables broadening of absorption spectrum in
  solar cells},\ }\href@noop {} {\bibfield  {journal} {\bibinfo  {journal}
  {Journal of the American Chemical Society}\ }\textbf {\bibinfo {volume}
  {136}},\ \bibinfo {pages} {8094} (\bibinfo {year}
  {2014}{\natexlab{a}})}\BibitemShut {NoStop}%
\bibitem [{\citenamefont {Ogomi}\ \emph {et~al.}(2014)\citenamefont {Ogomi},
  \citenamefont {Morita}, \citenamefont {Tsukamoto}, \citenamefont {Saitho},
  \citenamefont {Fujikawa}, \citenamefont {Shen}, \citenamefont {Toyoda},
  \citenamefont {Yoshino}, \citenamefont {Pandey}, \citenamefont {Ma} \emph
  {et~al.}}]{ref27}%
  \BibitemOpen
  \bibfield  {author} {\bibinfo {author} {\bibfnamefont {Y.}~\bibnamefont
  {Ogomi}}, \bibinfo {author} {\bibfnamefont {A.}~\bibnamefont {Morita}},
  \bibinfo {author} {\bibfnamefont {S.}~\bibnamefont {Tsukamoto}}, \bibinfo
  {author} {\bibfnamefont {T.}~\bibnamefont {Saitho}}, \bibinfo {author}
  {\bibfnamefont {N.}~\bibnamefont {Fujikawa}}, \bibinfo {author}
  {\bibfnamefont {Q.}~\bibnamefont {Shen}}, \bibinfo {author} {\bibfnamefont
  {T.}~\bibnamefont {Toyoda}}, \bibinfo {author} {\bibfnamefont
  {K.}~\bibnamefont {Yoshino}}, \bibinfo {author} {\bibfnamefont {S.~S.}\
  \bibnamefont {Pandey}}, \bibinfo {author} {\bibfnamefont {T.}~\bibnamefont
  {Ma}}, \emph {et~al.},\ }\bibfield  {title} {\bibinfo {title}
  {\ch{CH_3NH_3Sn_xPb_{(1--x)}I_3} perovskite solar cells covering up to 1060
  nm},\ }\href@noop {} {\bibfield  {journal} {\bibinfo  {journal} {The journal
  of physical chemistry letters}\ }\textbf {\bibinfo {volume} {5}},\ \bibinfo
  {pages} {1004} (\bibinfo {year} {2014})}\BibitemShut {NoStop}%
\bibitem [{\citenamefont {Mayengbam}\ \emph {et~al.}(2018)\citenamefont
  {Mayengbam}, \citenamefont {Tripathy},\ and\ \citenamefont {Palai}}]{ref28}%
  \BibitemOpen
  \bibfield  {author} {\bibinfo {author} {\bibfnamefont {R.}~\bibnamefont
  {Mayengbam}}, \bibinfo {author} {\bibfnamefont {S.}~\bibnamefont
  {Tripathy}},\ and\ \bibinfo {author} {\bibfnamefont {G.}~\bibnamefont
  {Palai}},\ }\bibfield  {title} {\bibinfo {title} {First-principle insights of
  electronic and optical properties of cubic organic--inorganic
  \ch{MAGe_xPb_{(1--x)}I_3} perovskites for photovoltaic applications},\
  }\href@noop {} {\bibfield  {journal} {\bibinfo  {journal} {The Journal of
  Physical Chemistry C}\ }\textbf {\bibinfo {volume} {122}},\ \bibinfo {pages}
  {28245} (\bibinfo {year} {2018})}\BibitemShut {NoStop}%
\bibitem [{\citenamefont {Ju}\ \emph {et~al.}(2017)\citenamefont {Ju},
  \citenamefont {Dai}, \citenamefont {Ma},\ and\ \citenamefont {Zeng}}]{ref31}%
  \BibitemOpen
  \bibfield  {author} {\bibinfo {author} {\bibfnamefont {M.-G.}\ \bibnamefont
  {Ju}}, \bibinfo {author} {\bibfnamefont {J.}~\bibnamefont {Dai}}, \bibinfo
  {author} {\bibfnamefont {L.}~\bibnamefont {Ma}},\ and\ \bibinfo {author}
  {\bibfnamefont {X.~C.}\ \bibnamefont {Zeng}},\ }\bibfield  {title} {\bibinfo
  {title} {Lead-free mixed tin and germanium perovskites for photovoltaic
  application},\ }\href@noop {} {\bibfield  {journal} {\bibinfo  {journal}
  {Journal of the American Chemical Society}\ }\textbf {\bibinfo {volume}
  {139}},\ \bibinfo {pages} {8038} (\bibinfo {year} {2017})}\BibitemShut
  {NoStop}%
\bibitem [{\citenamefont {Fang}\ \emph {et~al.}(2019)\citenamefont {Fang},
  \citenamefont {Shang}, \citenamefont {Hou}, \citenamefont {Zheng},
  \citenamefont {Du}, \citenamefont {Yang}, \citenamefont {Chou}, \citenamefont
  {Yang}, \citenamefont {Wang},\ and\ \citenamefont {Yang}}]{ref53}%
  \BibitemOpen
  \bibfield  {author} {\bibinfo {author} {\bibfnamefont {Z.}~\bibnamefont
  {Fang}}, \bibinfo {author} {\bibfnamefont {M.}~\bibnamefont {Shang}},
  \bibinfo {author} {\bibfnamefont {X.}~\bibnamefont {Hou}}, \bibinfo {author}
  {\bibfnamefont {Y.}~\bibnamefont {Zheng}}, \bibinfo {author} {\bibfnamefont
  {Z.}~\bibnamefont {Du}}, \bibinfo {author} {\bibfnamefont {Z.}~\bibnamefont
  {Yang}}, \bibinfo {author} {\bibfnamefont {K.-C.}\ \bibnamefont {Chou}},
  \bibinfo {author} {\bibfnamefont {W.}~\bibnamefont {Yang}}, \bibinfo {author}
  {\bibfnamefont {Z.~L.}\ \bibnamefont {Wang}},\ and\ \bibinfo {author}
  {\bibfnamefont {Y.}~\bibnamefont {Yang}},\ }\bibfield  {title} {\bibinfo
  {title} {Bandgap alignment of $\alpha$-\ch{CsPbI_3} perovskites with
  synergistically enhanced stability and optical performance via b-site minor
  doping},\ }\href@noop {} {\bibfield  {journal} {\bibinfo  {journal} {Nano
  Energy}\ }\textbf {\bibinfo {volume} {61}},\ \bibinfo {pages} {389} (\bibinfo
  {year} {2019})}\BibitemShut {NoStop}%
\bibitem [{\citenamefont {Wang}\ \emph {et~al.}(2020)\citenamefont {Wang},
  \citenamefont {Lei}, \citenamefont {Liu}, \citenamefont {He},\ and\
  \citenamefont {Zhang}}]{ref52}%
  \BibitemOpen
  \bibfield  {author} {\bibinfo {author} {\bibfnamefont {G.}~\bibnamefont
  {Wang}}, \bibinfo {author} {\bibfnamefont {M.}~\bibnamefont {Lei}}, \bibinfo
  {author} {\bibfnamefont {J.}~\bibnamefont {Liu}}, \bibinfo {author}
  {\bibfnamefont {Q.}~\bibnamefont {He}},\ and\ \bibinfo {author}
  {\bibfnamefont {W.}~\bibnamefont {Zhang}},\ }\bibfield  {title} {\bibinfo
  {title} {Improving the stability and optoelectronic properties of all
  inorganic less-pb perovskites by b-site doping for high-performance inorganic
  perovskite solar cells},\ }\href@noop {} {\bibfield  {journal} {\bibinfo
  {journal} {Solar RRL}\ }\textbf {\bibinfo {volume} {4}},\ \bibinfo {pages}
  {2000528} (\bibinfo {year} {2020})}\BibitemShut {NoStop}%
\bibitem [{\citenamefont {Kogo}\ \emph {et~al.}(2022)\citenamefont {Kogo},
  \citenamefont {Yamamoto},\ and\ \citenamefont {MURAKAMI}}]{ref51}%
  \BibitemOpen
  \bibfield  {author} {\bibinfo {author} {\bibfnamefont {A.}~\bibnamefont
  {Kogo}}, \bibinfo {author} {\bibfnamefont {K.}~\bibnamefont {Yamamoto}},\
  and\ \bibinfo {author} {\bibfnamefont {T.~N.}\ \bibnamefont {MURAKAMI}},\
  }\bibfield  {title} {\bibinfo {title} {Germanium ion doping of \ch{CsPbI_3}
  to obtain inorganic perovskite solar cells with low temperature processing},\
  }\href {http://iopscience.iop.org/article/10.35848/1347-4065/ac4927}
  {\bibfield  {journal} {\bibinfo  {journal} {Japanese Journal of Applied
  Physics}\ } (\bibinfo {year} {2022})}\BibitemShut {NoStop}%
\bibitem [{\citenamefont {Blaha}\ \emph {et~al.}(2001)\citenamefont {Blaha},
  \citenamefont {Schwarz}, \citenamefont {Madsen}, \citenamefont {Kvasnicka},
  \citenamefont {Luitz} \emph {et~al.}}]{ref32}%
  \BibitemOpen
  \bibfield  {author} {\bibinfo {author} {\bibfnamefont {P.}~\bibnamefont
  {Blaha}}, \bibinfo {author} {\bibfnamefont {K.}~\bibnamefont {Schwarz}},
  \bibinfo {author} {\bibfnamefont {G.~K.}\ \bibnamefont {Madsen}}, \bibinfo
  {author} {\bibfnamefont {D.}~\bibnamefont {Kvasnicka}}, \bibinfo {author}
  {\bibfnamefont {J.}~\bibnamefont {Luitz}}, \emph {et~al.},\ }\bibfield
  {title} {\bibinfo {title} {wien2k},\ }\href@noop {} {\bibfield  {journal}
  {\bibinfo  {journal} {An augmented plane wave+ local orbitals program for
  calculating crystal properties}\ } (\bibinfo {year} {2001})}\BibitemShut
  {NoStop}%
\bibitem [{\citenamefont {Perdew}\ \emph {et~al.}(1996)\citenamefont {Perdew},
  \citenamefont {Burke},\ and\ \citenamefont {Ernzerhof}}]{ref34}%
  \BibitemOpen
  \bibfield  {author} {\bibinfo {author} {\bibfnamefont {J.~P.}\ \bibnamefont
  {Perdew}}, \bibinfo {author} {\bibfnamefont {K.}~\bibnamefont {Burke}},\ and\
  \bibinfo {author} {\bibfnamefont {M.}~\bibnamefont {Ernzerhof}},\ }\bibfield
  {title} {\bibinfo {title} {Generalized gradient approximation made simple},\
  }\href@noop {} {\bibfield  {journal} {\bibinfo  {journal} {\textit{Physical
  review letters}}\ }\textbf {\bibinfo {volume} {\textbf{77}}},\ \bibinfo
  {pages} {3865} (\bibinfo {year} {1996})}\BibitemShut {NoStop}%
\bibitem [{\citenamefont {Sun}\ \emph {et~al.}(2016)\citenamefont {Sun},
  \citenamefont {Li}, \citenamefont {Feng},\ and\ \citenamefont {Li}}]{ref35}%
  \BibitemOpen
  \bibfield  {author} {\bibinfo {author} {\bibfnamefont {P.-P.}\ \bibnamefont
  {Sun}}, \bibinfo {author} {\bibfnamefont {Q.-S.}\ \bibnamefont {Li}},
  \bibinfo {author} {\bibfnamefont {S.}~\bibnamefont {Feng}},\ and\ \bibinfo
  {author} {\bibfnamefont {Z.-S.}\ \bibnamefont {Li}},\ }\bibfield  {title}
  {\bibinfo {title} {Mixed \ch{Ge}/\ch{Pb} perovskite light absorbers with an
  ascendant efficiency explored from theoretical view},\ }\href@noop {}
  {\bibfield  {journal} {\bibinfo  {journal} {Physical Chemistry Chemical
  Physics}\ }\textbf {\bibinfo {volume} {18}},\ \bibinfo {pages} {14408}
  (\bibinfo {year} {2016})}\BibitemShut {NoStop}%
\bibitem [{\citenamefont {Qian}\ \emph {et~al.}(2016)\citenamefont {Qian},
  \citenamefont {Xu},\ and\ \citenamefont {Tian}}]{ref36}%
  \BibitemOpen
  \bibfield  {author} {\bibinfo {author} {\bibfnamefont {J.}~\bibnamefont
  {Qian}}, \bibinfo {author} {\bibfnamefont {B.}~\bibnamefont {Xu}},\ and\
  \bibinfo {author} {\bibfnamefont {W.}~\bibnamefont {Tian}},\ }\bibfield
  {title} {\bibinfo {title} {A comprehensive theoretical study of halide
  perovskites \ch{ABX_3}},\ }\href@noop {} {\bibfield  {journal} {\bibinfo
  {journal} {Organic Electronics}\ }\textbf {\bibinfo {volume} {37}},\ \bibinfo
  {pages} {61} (\bibinfo {year} {2016})}\BibitemShut {NoStop}%
\bibitem [{\citenamefont {Pazoki}\ \emph {et~al.}(2016)\citenamefont {Pazoki},
  \citenamefont {Jacobsson}, \citenamefont {Hagfeldt}, \citenamefont
  {Boschloo},\ and\ \citenamefont {Edvinsson}}]{ref37}%
  \BibitemOpen
  \bibfield  {author} {\bibinfo {author} {\bibfnamefont {M.}~\bibnamefont
  {Pazoki}}, \bibinfo {author} {\bibfnamefont {T.~J.}\ \bibnamefont
  {Jacobsson}}, \bibinfo {author} {\bibfnamefont {A.}~\bibnamefont {Hagfeldt}},
  \bibinfo {author} {\bibfnamefont {G.}~\bibnamefont {Boschloo}},\ and\
  \bibinfo {author} {\bibfnamefont {T.}~\bibnamefont {Edvinsson}},\ }\bibfield
  {title} {\bibinfo {title} {Effect of metal cation replacement on the
  electronic structure of metalorganic halide perovskites: Replacement of lead
  with alkaline-earth metals},\ }\href@noop {} {\bibfield  {journal} {\bibinfo
  {journal} {Physical Review B}\ }\textbf {\bibinfo {volume} {93}},\ \bibinfo
  {pages} {144105} (\bibinfo {year} {2016})}\BibitemShut {NoStop}%
\bibitem [{\citenamefont {Shirayama}\ \emph {et~al.}(2016)\citenamefont
  {Shirayama}, \citenamefont {Kadowaki}, \citenamefont {Miyadera},
  \citenamefont {Sugita}, \citenamefont {Tamakoshi}, \citenamefont {Kato},
  \citenamefont {Fujiseki}, \citenamefont {Murata}, \citenamefont {Hara},
  \citenamefont {Murakami} \emph {et~al.}}]{ref38}%
  \BibitemOpen
  \bibfield  {author} {\bibinfo {author} {\bibfnamefont {M.}~\bibnamefont
  {Shirayama}}, \bibinfo {author} {\bibfnamefont {H.}~\bibnamefont {Kadowaki}},
  \bibinfo {author} {\bibfnamefont {T.}~\bibnamefont {Miyadera}}, \bibinfo
  {author} {\bibfnamefont {T.}~\bibnamefont {Sugita}}, \bibinfo {author}
  {\bibfnamefont {M.}~\bibnamefont {Tamakoshi}}, \bibinfo {author}
  {\bibfnamefont {M.}~\bibnamefont {Kato}}, \bibinfo {author} {\bibfnamefont
  {T.}~\bibnamefont {Fujiseki}}, \bibinfo {author} {\bibfnamefont
  {D.}~\bibnamefont {Murata}}, \bibinfo {author} {\bibfnamefont
  {S.}~\bibnamefont {Hara}}, \bibinfo {author} {\bibfnamefont {T.~N.}\
  \bibnamefont {Murakami}}, \emph {et~al.},\ }\bibfield  {title} {\bibinfo
  {title} {Optical transitions in hybrid perovskite solar cells: ellipsometry,
  density functional theory, and quantum efficiency analyses for
  \ch{CH_3NH_3PbI_3}},\ }\href@noop {} {\bibfield  {journal} {\bibinfo
  {journal} {Physical Review Applied}\ }\textbf {\bibinfo {volume} {5}},\
  \bibinfo {pages} {014012} (\bibinfo {year} {2016})}\BibitemShut {NoStop}%
\bibitem [{\citenamefont {Nagane}\ \emph {et~al.}(2018)\citenamefont {Nagane},
  \citenamefont {Ghosh}, \citenamefont {Hoye}, \citenamefont {Zhao},
  \citenamefont {Ahmad}, \citenamefont {Walker}, \citenamefont {Islam},
  \citenamefont {Ogale},\ and\ \citenamefont {Sadhanala}}]{ref39}%
  \BibitemOpen
  \bibfield  {author} {\bibinfo {author} {\bibfnamefont {S.}~\bibnamefont
  {Nagane}}, \bibinfo {author} {\bibfnamefont {D.}~\bibnamefont {Ghosh}},
  \bibinfo {author} {\bibfnamefont {R.~L.}\ \bibnamefont {Hoye}}, \bibinfo
  {author} {\bibfnamefont {B.}~\bibnamefont {Zhao}}, \bibinfo {author}
  {\bibfnamefont {S.}~\bibnamefont {Ahmad}}, \bibinfo {author} {\bibfnamefont
  {A.~B.}\ \bibnamefont {Walker}}, \bibinfo {author} {\bibfnamefont {M.~S.}\
  \bibnamefont {Islam}}, \bibinfo {author} {\bibfnamefont {S.}~\bibnamefont
  {Ogale}},\ and\ \bibinfo {author} {\bibfnamefont {A.}~\bibnamefont
  {Sadhanala}},\ }\bibfield  {title} {\bibinfo {title} {Lead-free perovskite
  semiconductors based on germanium--tin solid solutions: structural and
  optoelectronic properties},\ }\href@noop {} {\bibfield  {journal} {\bibinfo
  {journal} {The Journal of Physical Chemistry C}\ }\textbf {\bibinfo {volume}
  {122}},\ \bibinfo {pages} {5940} (\bibinfo {year} {2018})}\BibitemShut
  {NoStop}%
\bibitem [{\citenamefont {Brivio}\ \emph {et~al.}(2014)\citenamefont {Brivio},
  \citenamefont {Butler}, \citenamefont {Walsh},\ and\ \citenamefont
  {Van~Schilfgaarde}}]{ref40}%
  \BibitemOpen
  \bibfield  {author} {\bibinfo {author} {\bibfnamefont {F.}~\bibnamefont
  {Brivio}}, \bibinfo {author} {\bibfnamefont {K.~T.}\ \bibnamefont {Butler}},
  \bibinfo {author} {\bibfnamefont {A.}~\bibnamefont {Walsh}},\ and\ \bibinfo
  {author} {\bibfnamefont {M.}~\bibnamefont {Van~Schilfgaarde}},\ }\bibfield
  {title} {\bibinfo {title} {Relativistic quasiparticle self-consistent
  electronic structure of hybrid halide perovskite photovoltaic absorbers},\
  }\href@noop {} {\bibfield  {journal} {\bibinfo  {journal} {Physical Review
  B}\ }\textbf {\bibinfo {volume} {89}},\ \bibinfo {pages} {155204} (\bibinfo
  {year} {2014})}\BibitemShut {NoStop}%
\bibitem [{\citenamefont {Tran}\ and\ \citenamefont {Blaha}(2009)}]{ref42}%
  \BibitemOpen
  \bibfield  {author} {\bibinfo {author} {\bibfnamefont {F.}~\bibnamefont
  {Tran}}\ and\ \bibinfo {author} {\bibfnamefont {P.}~\bibnamefont {Blaha}},\
  }\bibfield  {title} {\bibinfo {title} {Accurate band gaps of semiconductors
  and insulators with a semilocal exchange-correlation potential},\ }\href@noop
  {} {\bibfield  {journal} {\bibinfo  {journal} {Physical review letters}\
  }\textbf {\bibinfo {volume} {102}},\ \bibinfo {pages} {226401} (\bibinfo
  {year} {2009})}\BibitemShut {NoStop}%
\bibitem [{\citenamefont {Dixit}\ \emph {et~al.}(2011)\citenamefont {Dixit},
  \citenamefont {Tandon}, \citenamefont {Cottenier}, \citenamefont {Saniz},
  \citenamefont {Lamoen}, \citenamefont {Partoens}, \citenamefont
  {Van~Speybroeck},\ and\ \citenamefont {Waroquier}}]{ref48}%
  \BibitemOpen
  \bibfield  {author} {\bibinfo {author} {\bibfnamefont {H.}~\bibnamefont
  {Dixit}}, \bibinfo {author} {\bibfnamefont {N.}~\bibnamefont {Tandon}},
  \bibinfo {author} {\bibfnamefont {S.}~\bibnamefont {Cottenier}}, \bibinfo
  {author} {\bibfnamefont {R.}~\bibnamefont {Saniz}}, \bibinfo {author}
  {\bibfnamefont {D.}~\bibnamefont {Lamoen}}, \bibinfo {author} {\bibfnamefont
  {B.}~\bibnamefont {Partoens}}, \bibinfo {author} {\bibfnamefont
  {V.}~\bibnamefont {Van~Speybroeck}},\ and\ \bibinfo {author} {\bibfnamefont
  {M.}~\bibnamefont {Waroquier}},\ }\bibfield  {title} {\bibinfo {title}
  {Electronic structure and band gap of zinc spinel oxides beyond lda:
  \ch{ZnAl_2O_4}, \ch{ZnGa_2O_4} and \ch{ZnIn_2O_4}},\ }\href@noop {}
  {\bibfield  {journal} {\bibinfo  {journal} {New Journal of Physics}\ }\textbf
  {\bibinfo {volume} {13}},\ \bibinfo {pages} {063002} (\bibinfo {year}
  {2011})}\BibitemShut {NoStop}%
\bibitem [{\citenamefont {Traor{\'e}}\ \emph {et~al.}(2019)\citenamefont
  {Traor{\'e}}, \citenamefont {Bouder}, \citenamefont {Lafargue-Dit-Hauret},
  \citenamefont {Rocquefelte}, \citenamefont {Katan}, \citenamefont {Tran},\
  and\ \citenamefont {Kepenekian}}]{ref49}%
  \BibitemOpen
  \bibfield  {author} {\bibinfo {author} {\bibfnamefont {B.}~\bibnamefont
  {Traor{\'e}}}, \bibinfo {author} {\bibfnamefont {G.}~\bibnamefont {Bouder}},
  \bibinfo {author} {\bibfnamefont {W.}~\bibnamefont {Lafargue-Dit-Hauret}},
  \bibinfo {author} {\bibfnamefont {X.}~\bibnamefont {Rocquefelte}}, \bibinfo
  {author} {\bibfnamefont {C.}~\bibnamefont {Katan}}, \bibinfo {author}
  {\bibfnamefont {F.}~\bibnamefont {Tran}},\ and\ \bibinfo {author}
  {\bibfnamefont {M.}~\bibnamefont {Kepenekian}},\ }\bibfield  {title}
  {\bibinfo {title} {Efficient and accurate calculation of band gaps of halide
  perovskites with the tran-blaha modified becke-johnson potential},\
  }\href@noop {} {\bibfield  {journal} {\bibinfo  {journal} {Physical Review
  B}\ }\textbf {\bibinfo {volume} {99}},\ \bibinfo {pages} {035139} (\bibinfo
  {year} {2019})}\BibitemShut {NoStop}%
\bibitem [{\citenamefont {Murnaghan}(1937)}]{ref44}%
  \BibitemOpen
  \bibfield  {author} {\bibinfo {author} {\bibfnamefont {F.~D.}\ \bibnamefont
  {Murnaghan}},\ }\bibfield  {title} {\bibinfo {title} {Finite deformations of
  an elastic solid},\ }\href@noop {} {\bibfield  {journal} {\bibinfo  {journal}
  {American Journal of Mathematics}\ }\textbf {\bibinfo {volume} {59}},\
  \bibinfo {pages} {235} (\bibinfo {year} {1937})}\BibitemShut {NoStop}%
\bibitem [{\citenamefont {Hao}\ \emph {et~al.}(2014{\natexlab{b}})\citenamefont
  {Hao}, \citenamefont {Stoumpos}, \citenamefont {Chang},\ and\ \citenamefont
  {Kanatzidis}}]{ref45}%
  \BibitemOpen
  \bibfield  {author} {\bibinfo {author} {\bibfnamefont {F.}~\bibnamefont
  {Hao}}, \bibinfo {author} {\bibfnamefont {C.~C.}\ \bibnamefont {Stoumpos}},
  \bibinfo {author} {\bibfnamefont {R.~P.}\ \bibnamefont {Chang}},\ and\
  \bibinfo {author} {\bibfnamefont {M.~G.}\ \bibnamefont {Kanatzidis}},\
  }\bibfield  {title} {\bibinfo {title} {Anomalous band gap behavior in mixed
  \ch{Sn} and \ch{Pb} perovskites enables broadening of absorption spectrum in
  solar cells},\ }\href@noop {} {\bibfield  {journal} {\bibinfo  {journal}
  {Journal of the American Chemical Society}\ }\textbf {\bibinfo {volume}
  {136}},\ \bibinfo {pages} {8094} (\bibinfo {year}
  {2014}{\natexlab{b}})}\BibitemShut {NoStop}%
\end{thebibliography}%


\begin{thebibliography}{0}%
\makeatletter
\providecommand \@ifxundefined [1]{%
 \@ifx{#1\undefined}
}%
\providecommand \@ifnum [1]{%
 \ifnum #1\expandafter \@firstoftwo
 \else \expandafter \@secondoftwo
 \fi
}%
\providecommand \@ifx [1]{%
 \ifx #1\expandafter \@firstoftwo
 \else \expandafter \@secondoftwo
 \fi
}%
\providecommand \natexlab [1]{#1}%
\providecommand \enquote  [1]{``#1''}%
\providecommand \bibnamefont  [1]{#1}%
\providecommand \bibfnamefont [1]{#1}%
\providecommand \citenamefont [1]{#1}%
\providecommand \href@noop [0]{\@secondoftwo}%
\providecommand \href [0]{\begingroup \@sanitize@url \@href}%
\providecommand \@href[1]{\@@startlink{#1}\@@href}%
\providecommand \@@href[1]{\endgroup#1\@@endlink}%
\providecommand \@sanitize@url [0]{\catcode `\\12\catcode `\$12\catcode
  `\&12\catcode `\#12\catcode `\^12\catcode `\_12\catcode `\%12\relax}%
\providecommand \@@startlink[1]{}%
\providecommand \@@endlink[0]{}%
\providecommand \url  [0]{\begingroup\@sanitize@url \@url }%
\providecommand \@url [1]{\endgroup\@href {#1}{\urlprefix }}%
\providecommand \urlprefix  [0]{URL }%
\providecommand \Eprint [0]{\href }%
\providecommand \doibase [0]{https://doi.org/}%
\providecommand \selectlanguage [0]{\@gobble}%
\providecommand \bibinfo  [0]{\@secondoftwo}%
\providecommand \bibfield  [0]{\@secondoftwo}%
\providecommand \translation [1]{[#1]}%
\providecommand \BibitemOpen [0]{}%
\providecommand \bibitemStop [0]{}%
\providecommand \bibitemNoStop [0]{.\EOS\space}%
\providecommand \EOS [0]{\spacefactor3000\relax}%
\providecommand \BibitemShut  [1]{\csname bibitem#1\endcsname}%
\let\auto@bib@innerbib\@empty
\end{thebibliography}%

\end{document}